\documentclass{aa}  
\bibpunct{(}{)}{;}{a}{}{,} 
\usepackage[]{natbib}
\usepackage{graphicx}
\usepackage{txfonts}
\begin{document} 

   \title{Diversity of planetary systems in low-mass disks:} 
   \subtitle{Terrestrial-type planet formation and water delivery}

   \author{M. P. Ronco\thanks{mpronco@fcaglp.unlp.edu.ar}
          \and
          G. C. de El\'{\i}a}

\offprints{M. P. Ronco}

   \institute{Facultad de Ciencias Astron\'omicas y Geof\'\i sicas, Universidad 
Nacional de La Plata and Instituto de Astrof\'{\i}sica de La Plata, CCT La Plata-CONICET-UNLP,
   Paseo del Bosque S/N (1900), La Plata, Argentina}

   \date{Received / Accepted }

  \abstract
   {Several studies, observational and theoretical, suggest that 
   planetary systems with only rocky planets should be the most common in the Universe.} 
   {We study the diversity of planetary systems that could form around Sun-like stars in low-mass 
    disks without gas giants planets. Especially we focus on the formation process of terrestrial planets in the habitable zone
    and analyze their water contents with the goal to determine systems of astrobiological interest.
    Besides, we study the formation of planets on wide orbits because they are sensitive to be detected with the 
    microlensing technique.}
   {N-body simulations of high resolution are developed for a wide range of surface density profiles. A bimodal 
     distribution of planetesimals and planetary embryos with different physical and orbital configurations is
     used in order to perform the planetary accretion process. The surface density profile combines a power law to the inside of 
     the disk of the form $r^{-\gamma}$, with an exponential decay to the outside. We perform simulations adopting a disk of 
     $0.03M_\odot$ and values of $\gamma = 0.5$, $1$ and $1.5$.}
   {All our simulations form planets in the Habitable Zone (HZ) with different masses and final water contents depending on 
     the three different profiles. For $\gamma =$ 0.5, our simulations produce three planets in the HZ with masses ranging 
     from 0.03 $M_{\oplus}$ to 0.1 $M_{\oplus}$ and water contents between 0.2 and 16 Earth oceans 
     (1 Earth ocean $=2.8\times10^{-4}M_\oplus$). For $\gamma =$ 1, 
     three planets form in the HZ with masses between 0.18 $M_{\oplus}$ and 0.52 $M_{\oplus}$ and water contents from 34 to 
     167 Earth oceans. At last, for $\gamma =$ 1.5, we find four planets in the HZ with masses ranging from 0.66 $M_{\oplus}$
     to 2.21 $M_\oplus$ and water contents between 192 and 2326 Earth oceans. This profile shows distinctive results because
     is the only one of those studied here that leads to the formation of \emph{water worlds}.}
   {Since planetary systems with $\gamma = 1$ and $1.5$ present planets in the HZ with suitable masses to retain a long-live 
     atmosphere and to maintain plate tectonics, they seem to be the most outstanding candidates to be potentially habitable. 
     Particularly, these systems form Earths and Super-Earths of at least $3M_\oplus$ around the snow line which are sensitive 
     to be discovered by the microlensing technique.}

                             
   \maketitle

\section{Introduction}

The accretion process that leads to the formation of terrestrial planets
is strongly dependent on the mass distribution in the system and on the
presence of gas giant planets. From this, to analyze the diversity of
planetary systems that could form around solar-type stars, it is necessary
to consider protoplanetary disks with different surface density profiles
as well as several physical and orbital parameters for the gas giants.

During the last years, several observational works have
suggested that the planetary systems consisting only of rocky planets
would seem to be the most common in the Universe. In fact, using precise
radial velocity measurements from the Keck planet search, \citet{Cumming2008}
inferred that 17\%-19\% of the solar-type stars
have giant planets with masses $M > $ 100 $M_{\oplus}$ within 20 AU. More
recently, \citet{Mayor2012} analyzed the results of an
eight-year survey carried out at the La Silla Observatory with the HARPS
spectrograph and suggested that about 14\% of the solar-type stars have
planets with masses $M >$ 50 $M_{\oplus}$ within 5 AU. On the other hand, many 
theoretical works complement these results. For example,
\citet{Mordasini2009} developed a great number of
planet population synthesis calculations of solar-like stars within the
framework of the core accretion scenario. From this theoretical study,
these authors indicated that the occurrence rate of planets with masses
$M >$ 100 $M_{\oplus}$ is 14.3\%, which is in agreement with that obtained
by \citet{Cumming2008}. More recently, \citet{Miguel2011} developed a semi-analytical code for computing the
planetary system formation, based on the core instability model for the gas
accretion of the embryos and the oligarchic growth regime for the accretion
of the solid cores. The most important result obtained by such authors suggests
that those planetary systems with only small rocky planets represent probably
the vast majority in the Universe.

The standard ``model of solar nebula'' (MSN) developed by\citet{Weidenschilling1977}
and \citet{Hayashi1981} predicts that the surface densities of dust
materials and gases vary approximately as $r^{-1.5}$, where $r$ is the distance
from the Sun. This model is constructed by adding the solar complement of light
elements to each planet and then, by distributing such augmented mass uniformly
across an annulus around the location of each planet. \citet{Davis2005}
reanalyzed the model of solar nebula and predicted that the surface density follows
a decay rate of $r^{-0.5}$ in the inner region and a subsequent exponential decay. Later,
\citet{Desch2007} adopted the starting positions of the planets in the Nice
model \citep{Tsiganis2005} and suggested that the surface density of the
solar nebula varies approximately as $r^{-2.2}$. 
On the other hand, detailed models
of structure and evolution of protoplanetary disks \citep{Dullemond2007,Garaud2007} suggest that 
the surface density falls off with radius
much less steeply (as $r^{-1}$) than that assumed for the MSN. More recently,
\citet{Andrews2009,Andrews2010} analyzed protoplanetary disk structures in
Ophiuchus star-forming region. They inferred that the surface density profile follows a power
law in the inner disk of the form $r^{-\gamma}$ and an exponential taper at large radii, where
$\gamma$ ranges from 0.4 to 1.1 and shows a median value of 0.9.

Several authors have tried to build a minimum-mass extrasolar nebula (MMEN) using observational 
data of exoplanets. On the one hand, \citet{Kuchner2004} used 11 systems with Jupiter-mass planets detected by 
radial velocity to construct an MMEN. On the other hand, \citet{Chiang&Laughlin2013} used Kepler data concerning 
planetary candidates with radii R $<$ 5R to built an MMEN. Both analyses produced steep density profiles. In fact, 
\citet{Kuchner2004} predicted that the surface density varies as $r^{-2}$ for the gas giants, while \citet{Chiang&Laughlin2013}
suggested that the surface density follows a decay rate of $r^{-1.6}$ for the super-Earths. Recently, 
\citet{Raymond&Cossou2014} suggested that it is inconsistent to assume a universal disk profile. In fact, these authors 
predicted that the minimum-mass disks calculated from multiple-planet systems show a wide range of surface density slopes.

Several previous studies have analyzed the effects of the surface density profile on the 
terrestrial planet formation in a wide range of scenarios. On the one hand, \citet{Chambers&Cassen2002}  
and \citet{Raymond2005} examined the process of planetary accretion in disks with varying 
surface density profiles in presence of giant planets. On the other hand, \citet{Kokubo2006} 
investigated the final assemblage of terrestrial planets from different surface density profiles 
considering gas-free cases without gas giants. While these authors assumed disks with different masses, 
they only included planetary embryos (no planetesimals) in a narrow radial range of the system between 
0.5~AU – 1.5~AU. \citet{Raymond2007a} simulated the terrestrial planet formation without gas giants 
for a wide range of stellar masses. For Sun-like stars, they developed simulations in a wider radial 
range (from 0.5~AU to 4~AU) than that assumed by \citet{Kokubo2006}. However, like \citet{Kokubo2006}, 
\citet{Raymond2007a} did not examine the effects of a planetesimal population in their simulations.

Here, we show results of N-body simulations aimed at analyzing the process of
formation of terrestrial planets and water delivery in absence of gas giants. It is important to 
highlight that these are high resolution simulations that include planetary embryos and planetesimals.
In particular, our work focuses on low-mass protoplanetary disks for a wide range
of surface density profiles. This study is motivated by an interesting result
obtained by \citet{Miguel2011}, which indicates that a planetary
system composed by only rocky planets is the most common outcome obtained from a
low-mass disk (namely, $\lesssim$ 0.03 $M_{\odot}$) for different surface density
profiles. The most important goal of the present work is to analyze the potential
habitability of the terrestrial planets formed in our simulations. From this, we
will be able to determine if the planetary systems under consideration result to
be targets of astrobiological interest.

Basically, the permanent presence of liquid water on the surface of a planet is the
main condition required for habitability. However, the existence of liquid water
is a necessary but not sufficient condition for planetary habitability. In fact,
the existence of organic material, the preservation of a suitable atmosphere, the
presence of a magnetic field, and the tectonic activity represent other relevant conditions
for the habitability of a planet. The N-body simulations presented here allow us to describe
the dynamical evolution and the accretion history of a planetary system. From this, it
is possible to analyze the water delivery to the resulting planets, primarily to those formed
in the habitable zone (HZ), which is defined as the circumstellar region inside which a planet
can retain liquid water on its surface. From this, the criteria adopted in the present paper
to determine the potential habitability of a planet will be based on its location in the system
and its final water content.

This paper is therefore structured as follows. In Sect. 2, we present the main properties of
the protoplanetary disks used in our simulations. Then, we discuss the main characteristics
of the N-body code and outline our choice of initial conditions in Sect. 3. In Sect. 4, we
present the results of all simulations. Finally, we discuss such results within the framework
of current knowledge of planetary systems and present our conclusions in Sect. 6.  

 \section{Protoplanetary Disk: Properties}
Here we describe the model of the protoplanetary disk and then define 
the parameters we need to develop our simulations.

The surface density profile is one of the most relevant parameters to determine the distribution of material 
in the disk. Particularly, in this model, the gas surface density profile that represents the structure of the protoplanetary 
disk is given by
\begin{eqnarray}
\Sigma_{\textrm{g}}(r) = \Sigma^{0}_{\textrm{g}}\left(\dfrac{r}{r_{\textrm{c}}}\right)^{-\gamma}e^{-(\frac{r}{r_{\textrm{c}}})^{2-\gamma}},
\label{eq:gas}
\end{eqnarray}
where $\Sigma_{\textrm{g}}^{0}$ is a normalization constant, $r_{\textrm{c}}$ a characteristic radius and $\gamma$ the 
exponent that represents the surface density gradient. By integrating Eq. \ref{eq:gas} over the total area of the disk, 
$\Sigma_{\textrm{g}}^{0}$ can be expressed by

\begin{eqnarray}
\Sigma^{0}_{\textrm{g}} = (2-\gamma)\dfrac{M_{\textrm{d}}}{2\pi R^{2}_{\textrm{c}}},
\label{eq:gas2}
\end{eqnarray} 

This surface density profile that combines a power law to the inside of the disk, with an exponential decay to 
the outside, is based on the similarity solutions of the surface
density of a thin Keplerian disc subject to the gravity of a point mass ($M_\star$) central star 
\citep{Lynden-Bell1974,Hartmann1998}.

Analogously, the solid surface density profile $\Sigma_{\textrm{s}}(r)$ is represented by
\begin{eqnarray}
\Sigma_{\textrm{s}}(r) = \Sigma^{0}_{\textrm{s}}\eta_{\textrm{ice}}\left(\dfrac{r}{r_{\textrm{c}}}\right)^{-\gamma}e^{-(\frac{r}{r_{\textrm{c}}})^{2-\gamma}},
\label{eq:solidos}
\end{eqnarray}
where $\eta_{\textrm{ice}}$ represents an increase in the amount of solid material due to the condensation of water 
beyond the snow line. According to the MSN of 
\citet{Hayashi1981}, $\eta_{\textrm{ice}}$ is $1/4$ inside and $1$ outside the snow line which is located at 2.7~AU \footnote{
Although we use the classic model of \citet{Hayashi1981}, is worth noting that \citet{Lodders2003} found a much lower value for
the increase in the amount of solids due to water condensation beyond the snow line. This value corresponds to a factor of 2 
instead of 4.}. 

Once we derived the value of $\Sigma^{0}_{\textrm{g}}$  the relation between both 
profiles allows us to know the abundance of heavy elements, which is given by
\begin{eqnarray}
\left(\dfrac{\Sigma^{0}_{\textrm{s}}}{\Sigma^{0}_{\textrm{g}}}\right)_{\star} = \left(\dfrac{\Sigma^{0}_{\textrm{s}}}{\Sigma^{0}_{\textrm{g}}}\right)_{\odot}10^{[\textrm{Fe}/\textrm{H}]} = z_{0}10^{[\textrm{Fe}/\textrm{H}]},
\end{eqnarray}
where $z_0$ is the primordial abundance of heavy elements in the Sun and $[\textrm{Fe}/\textrm{H}]$ the metallicity. 

After the model is presented, we need to quantify some parameters for the disk. We assume that the central star of the
protoplanetary disk is a solar metallicity star, $[\textrm{Fe}/\textrm{H}]=0$, with $M_\star = 1M_\odot$. Then, $\Sigma^{0}_{\textrm{s}} = z_{0}\Sigma^{0}_{\textrm{g}}$ 
where $z_0 = 0.0149$ \citep{Lodders2003}. 
We consider a low-mass disk with $M_{\textrm{d}} = 0.03M_\odot$ because this value for the mass guarantee that the formation 
of giant planets is not performed \citep{Miguel2011}.

The exponent $\gamma$ is another relevant parameter for this model because it establishes an important characteristic of the
simulation scenario: the larger the exponent, more massive the disk in the inner part of it. In this work
we explore three different values for the exponent: $\gamma = 0.5$, $1$ and $1.5$. These values represent disks from rather 
flat ones ($\gamma = 0.5$) where the density of gas and solids are well distributed and there are no preferential 
areas for the accumulation of gas and solids, to steeper profiles ($\gamma = 1.5$) where there is an accumulation of gas and solids in 
the inner part of the disk and around the snow line.  
Finally, for the characteristic radius we adopt $r_{\textrm{c}} = 50$~AU, 
which represents a characteristic value of the different disk's observations studied by \citet{Andrews2010}. 

Regarding water content in the disk, we assume that the protoplanetary disk presents a radial compositional gradient.
Then we adopt an initial distribution of water similar to that used by \citet{Raymond2004, Raymond2006} and \citet{Mandell2007},
which is based on the one predicted by \citet{Abe2000}. Thus, the water 
content by mass $W(r)$ assumed as a function of the radial distance $r$ is given by:
\begin{eqnarray*}
W(r) =\left\{ \begin{array}{ll}
0.001\%,~~~~& r < 2~\textrm{AU}\\ 
0.1\%,~~~~ & 2~\textrm{AU} < r < 2.5~\textrm{AU}\\
5\%,~~~~ & 2.5~\textrm{AU} < r < 2.7~\textrm{AU}\\
50\%,~~~~ & r > 2.7~\textrm{AU}.
\end{array} \right.
\label{eq:agua}
\end{eqnarray*}
We assign this water distribution to each body in our simulations, based on its 
starting location. In particular, the model does not consider water loss during impacts and therefore the water
content represents an upper limit. Because this distribution is based on information about the Solar System, 
it is unknown yet if this is representative of the vast diversity of planetary systems in the Universe. However, 
we adopt it to study the water content and hence, the potential habitability of the resulting terrestrial planets. 

\section{N-body Method: Characteristics and Initial Conditions}

The initial time for our simulations represents the epoch in which the gas of the disk 
has already dissipated \footnote{ In the present work, we study the processes of planetary formation 
considering gas-free cases. Thus, we do not model the effects of gas on the planetesimals and planetary embryos 
of our systems. In particular, the type I migration \citep{Ward1997}, which leads to the orbital decay of embryos 
and planet-sized bodies through tidal interaction with the gaseous disk, could play a significant role in the 
evolution of these planetary systems. However, many quantitative aspects of the type I migration are still 
uncertain and so, we decide to neglect its effects in the present work. A detailed analysis concerning the 
action of the type I migration on the planetary systems of our simulations will be developed in a future study.}.

The numerical code used in our N-body simulations is the MERCURY code developed by \citet{Chambers1999}.
We particularly adopt the hybrid integrator which uses a second-order mixed variable symplectic algorithm to treat the 
interaction between objects with separations greater than 3 Hill radii, and a Burlisch-Stoer method for resolving 
closer encounters. In order to avoid any numerical error for 
small-perihelion orbits, we use a non-realistic size for the Sun's radius of 0.1~AU \citep{Raymond2009}. 

Since our main goal is to form planetary systems with terrestrial planets, we focus our study in the inner part 
of the protoplanetary disk, between 0.5~AU and 5~AU. The solid component of the disk presents a bimodal distribution 
formed by protoplanetary embryos and planetesimals. We use 1000 planetesimals in each simulation. Then, the number 
of protoplanetary embryos depends on each density profile. Collisions between both components are treated as 
inelastic mergers that conserve mass and water content. Since N-body simulations are very costly numerically and in order 
to reduce CPU time, the model considers gravitational interactions between embryos and planetesimals but not between 
planetesimals \citep{Raymond2006}. 
It is important to emphasize that these N-body high resolution simulations allow
us to describe in detail the dynamical processes involved during the formation and post evolution stages. 

With the total mass of the disk, $M_{\textrm{d}} = 0.03M_\odot$, we 
calculate the mass of solids in the study region and obtain the solid mass before and after the snow line for each profile.
For disks with $\gamma = 0.5$ the total mass of solids is $3.21M_\oplus$. Disks with $\gamma = 1$ present a total mass
of solids of $7.92M_\oplus$ and finally, disks with $\gamma = 1.5$ have $13.66M_\oplus$.  We then 
distribute the solid mass in accordance with various planetary accretion studies such as \citet{Kokubo1998}: 
each component, embryos and planetesimals, have half the total mass in the study region.
To distinguish both kinds of bodies, the mass adopted for the planetesimals is approximately an order 
of magnitude less than those associated with protoplanetary embryos. This is considered both for the \emph{inner zone}, 
between 0.5~AU and the snow line, and for the \emph{outer zone}, between the snow line and 5~AU. 

Since terrestrial planets in our Solar System 
could have formed in $100~$Myr~-~$200~$Myr \citep{Touboul2007, Dauphas2011} 
we integrate each simulation for at least $200$~Myr. To compute the inner orbit with 
enough precision we use a 6 day time step. Besides, each simulation conserved energy to at least one part in $10^3$.


In order to begin our simulations with the MERCURY code, we need to specify some important initial physical and orbital 
parameters. 
For disks with $\gamma = 0.5$ we use 24 embryos, 13 in the inner zone with masses of $0.02M_\oplus$ and 11 in the 
outer zone with masses of $0.13M_\oplus$. Planetesimals present masses of $3.87\times10^{-4}M_\oplus$ and 
$3.09\times10^{-3}M_\oplus$ in the inner and outer zone, respectively. For $\gamma = 1$ we place 30 embryos in the disk, 
20 in the inner zone with masses of $0.04M_\oplus$ and 10 in the outer zone with masses of $0.32M_\oplus$. For this density 
profile, planetesimals present masses of $1.18\times10^{-3}M_\oplus$ and $9.49\times10^{-3}M_\oplus$ in the inner and outer zone, 
respectively.
At last, disks with $\gamma = 1.5$ present 45 embryos, 35 in the inner zone with masses of $0.06M_\oplus$ and 10 in the 
outer zone with masses of $0.47M_\oplus$. Planetesimals in these protoplanetary disks have masses of $2.68\times10^{-3}M_\oplus$ 
and $0.021M_\oplus$ in the inner and outer zone, respectively. As we have mentioned, we use 1000 planetesimals in each 
simulation which are distributed between 0.5~AU and 5~AU, in order to efficiently model the effects of the dynamical friction. 
For any disk, physical densities of all planetesimals and protoplanetary embryos are assumed as 3 g~cm$^{-3}$ or 
1.5g~cm$^{-3}$ depending on whether they initially locate in the inner zone or in the outer zone, respectively. 

The orbital parameters, like initial eccentricities and inclinations, take random values less than 0.02 and $0.5^{\circ}$, 
respectively, both for embryos and planetesimals. In the same way, we adopt random values for the argument of pericenter 
$\omega$, longitude of ascending node $\Omega$ and the mean anomaly $M$ between $0^{\circ}$ and $360^{\circ}$.  
Finally, the semimayor axis of embryos and planetesimals are generated using the acceptance-rejection method developed 
by Jonh von Newmann. This technique indicates that if a number $a$ is selected randomly from the domain of a function $f$, 
and another number $f^{*}$ is given at random from the range of such function, so the condition $f^{*} \leq f(a)$ will 
generate a distribution for $a$ whose density is $f(a)da$. In our case, for each density profile under consideration, 
the $f$ function is represented by:
\begin{eqnarray}
f(a) = \dfrac{2\pi a\Sigma_{\textrm{s}}(a)} {M},
\end{eqnarray}
where $M$ represents the total mass of solids in the study region and $\Sigma_{\textrm{s}}(a)$ is
the solid density profile that we are using. Thus, the $a$ values obtained 
from this function will be accepted as initial conditions for the semimajor axes of embryos and planetesimals. 

\section{Results}

In this section we present results of the N-body simulations for the formation of terrestrial 
planets in low-mass disks, for different density profiles.

Given the stochastic nature of the accretion process, we perform three simulations for each value of $\gamma$ 
with different random number seeds.
Although the data analysis takes into account all the simulations per density profile, here
we show graphics of the most representative simulation.

The general purpose of this work is to analyze the diversity of planetary systems that we can perform with
the N-body simulations. From this, a particular goal is to determine if those terrestrial planets in the
habitable zone (HZ) are potentially habitable. 

Here we define the HZ for a solar-type star between 
0.8~AU and 1.5~AU in agreement with \citet{Kasting1993} and \citet{Selsis2007}. 
However, the evolution of Earth-like planets is a very complex process and locate a planet in the HZ does
not guarantee that there may developed life. For example, planets with very eccentric orbits may pass most of their
periods outside the HZ, not allowing long times of permanence of liquid water on their surfaces. In
order to avoid this problem we consider a planet is in the HZ and can hold liquid water if it has a perihelion
$\textrm{q} \geq 0.8~$AU and a aphelion $\textrm{Q} \leq 1.5~$AU \citep{deElia2013}. These criteria allow us
to distinguish potentially habitable planets.

Regarding water contents, we consider it significant
if it is comparable to that of Earth. The total amount of Earth's water is still uncertain because the mass of 
water in the actual mantle in not very well known yet. While the mass of water in the Earth's surface is $2.8 \times
10^{-4}M_\oplus$ (1 Earth ocean) the mass of water in the Earth's mantle is estimated to be between $0.8\times 10^{-4}M_\oplus$ and 
$8\times 10^{-4}M_\oplus$ \citep{Lecuyer1998}. On the other hand, \citet{Marty2012} suggested 
that the current water content in Earth's mantle is approximately $2\times 10^{-3}M_\oplus$. Following these studies 
the present-day water content in Earth should be $\sim 0.1\%$ to $0.2\%$ by mass. However, Earth should have had a 
greater amount of water in the early stages of its formation which has been lost during the process of the core's 
creation and due to consecutive impacts. In particular, as we consider inelastic collisions in our N-body simulations, 
the mass and water content are conserved and so, the final water contents in the formed planets represent upper limits.

\subsection{Simulations with $\gamma = 0.5$}

For this density profile we integrate three different simulations for 250~Myr \footnote{For this profile we did not 
extend over the simulations because the CPU time required is very high.}.

In general terms, the most important
characteristics of the systems can be described as follows. The entire study region, from 0.5~AU to 5~AU, contains at
the end of the simulation between 6 and 7 planets with a total mass ranging between $1.5M_\oplus$ and 
$1.6M_\oplus$. The mass of each planet is between $0.03M_{\oplus}$ and $0.57M_{\oplus}$. All simulations form planets in the
HZ and they seem to be very small. Their masses range from $0.03M_{\oplus}$ to $0.1M_{\oplus}$ and their water contents 
from $0.06\%$ to $5.37\%$ respect the total mass which represents from 0.2 to $\sim$ 6 Earth's oceans. 

\begin{figure*}[t]
 \centering
 \includegraphics[angle=0, width= 0.45\textwidth]{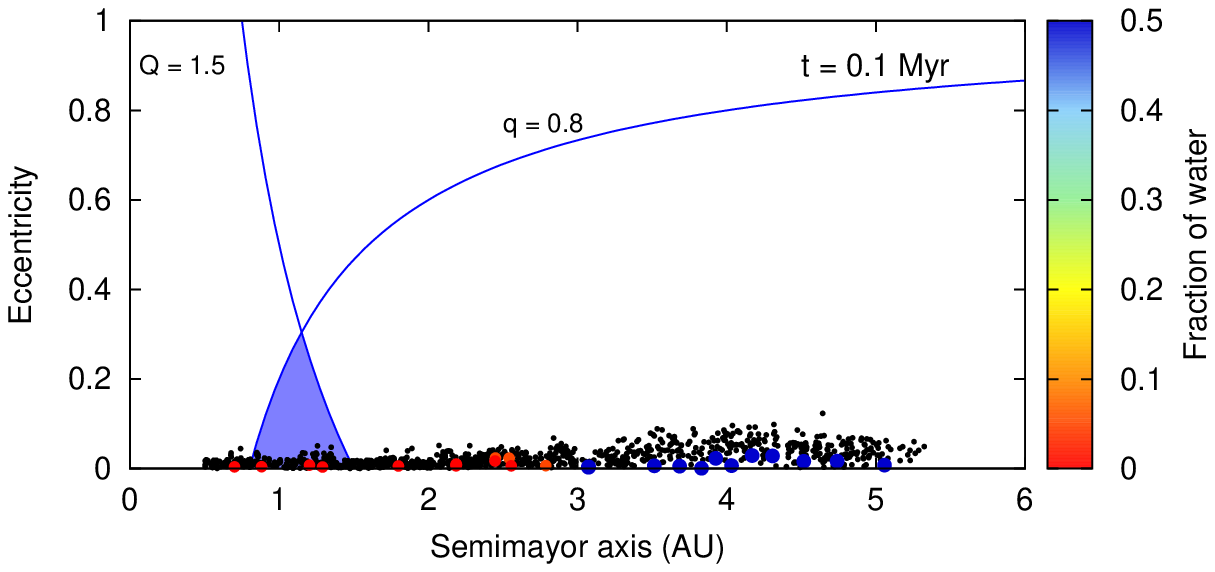}
 \includegraphics[angle=0, width= 0.45\textwidth]{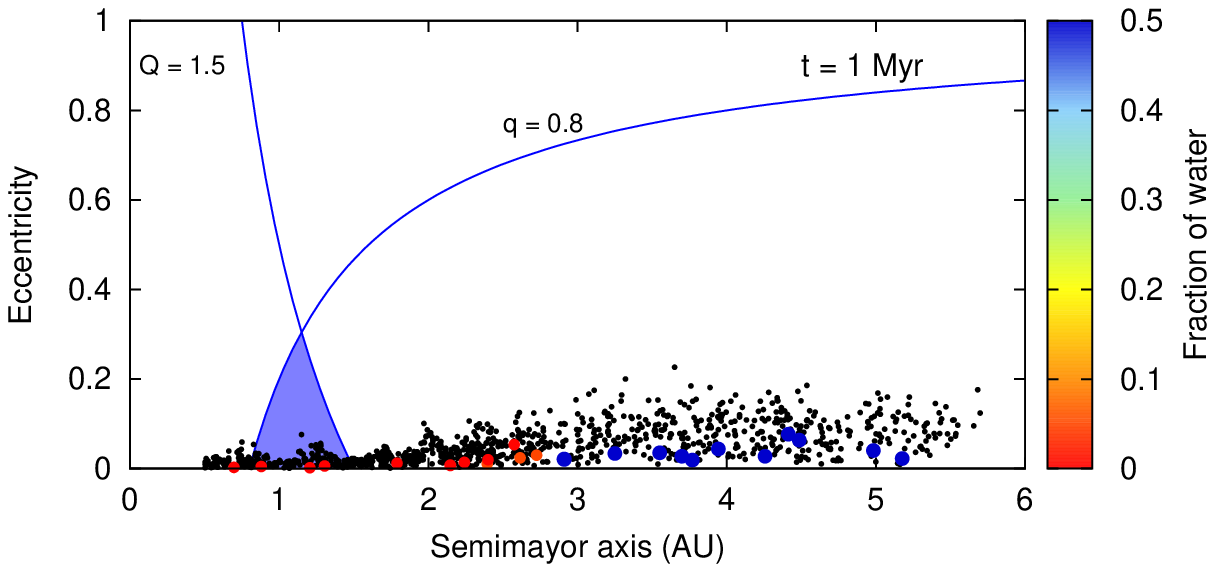}\\
 \includegraphics[angle=0, width= 0.45\textwidth]{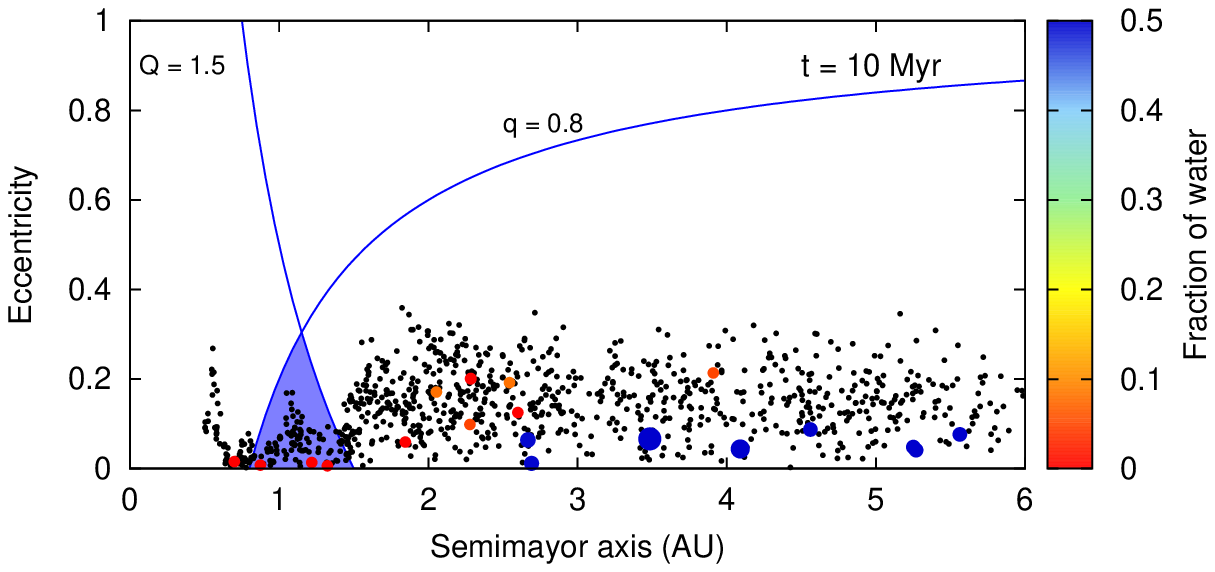}
 \includegraphics[angle=0, width= 0.45\textwidth]{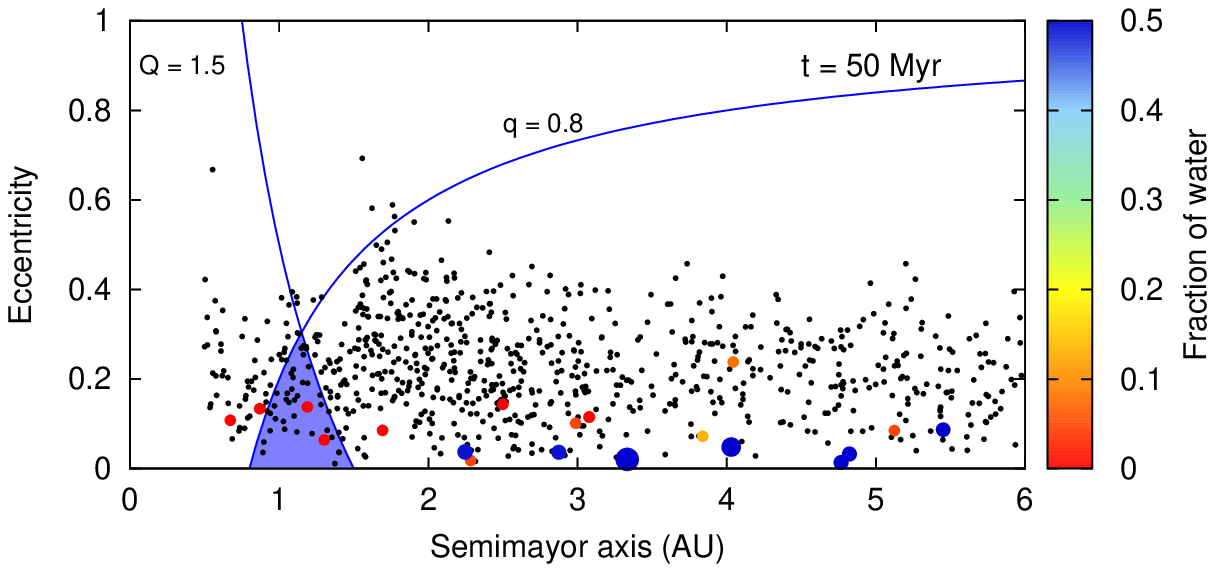}\\
 \includegraphics[angle=0, width= 0.45\textwidth]{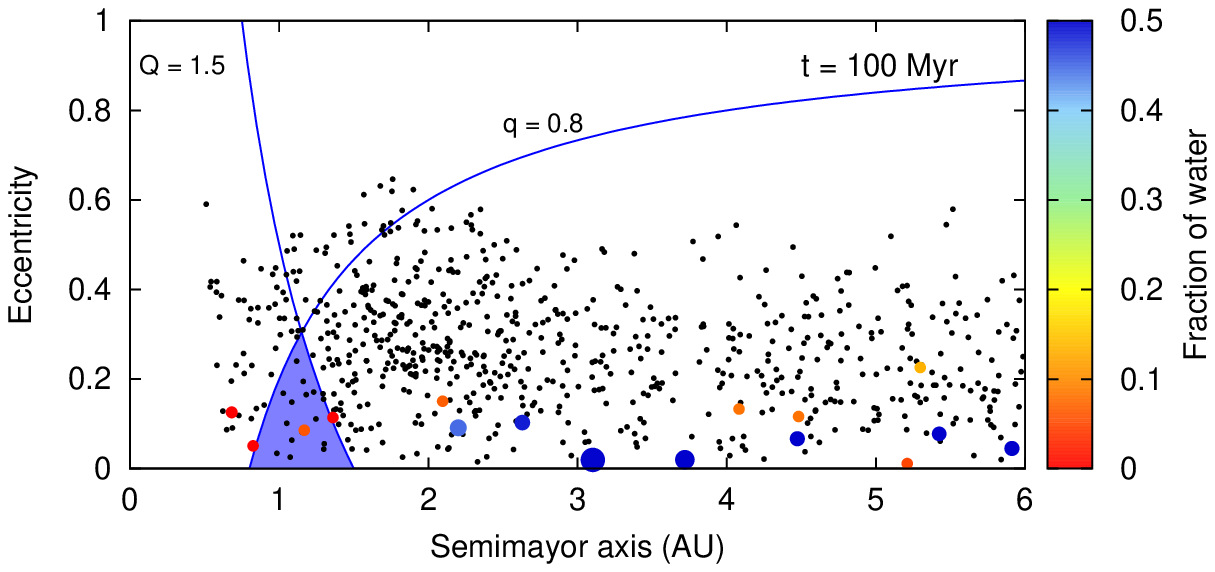}
 \includegraphics[angle=0, width= 0.45\textwidth]{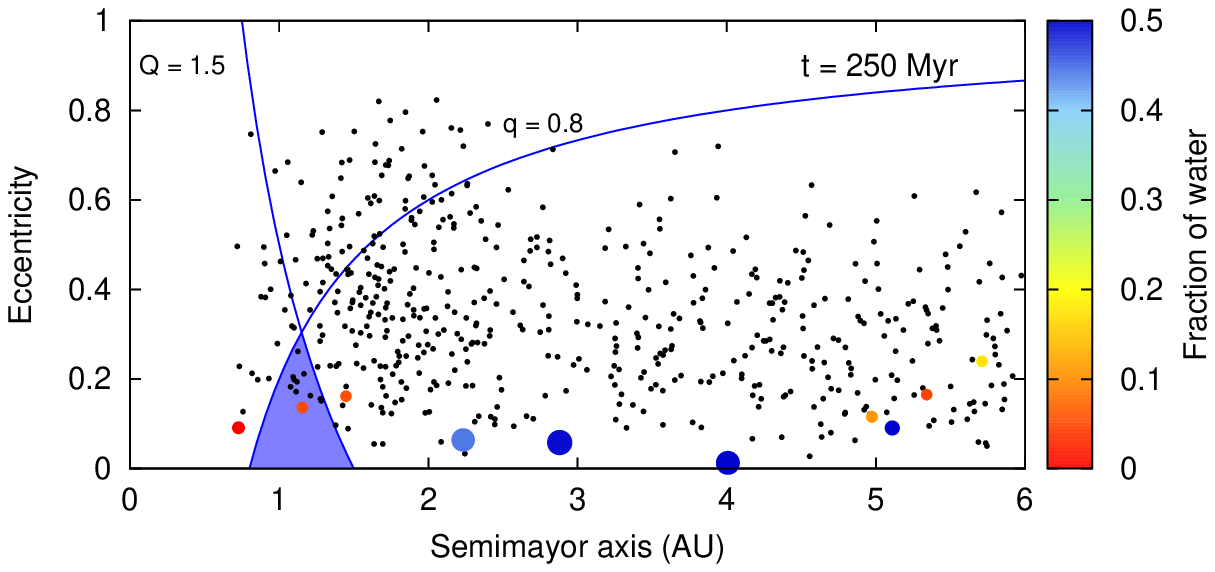}
 \caption{Evolution in time of the S3 simulation for $\gamma = 0.5$. The light-blue shaded
area represents the HZ and the curves with $q = 0.8$~AU and $Q = 1.5~$AU represent curves of constant perihelion and
aphelion, respectively. Planetary embryos are plotted as colored circles and planetesimals are plotted with black dots. 
The color scale represents the fraction of water of the embryos relative to their masses. In 
this case there is only one embryo in the HZ, located at 1.15~AU with a mass of $0.03M_\oplus$ and a water 
content of $5.37\%$ by mass which represents 6 Earth's oceans. Color figure is only available in the electronic 
version.} 
 \label{fig:g05-1}
\end{figure*}

In particular, Fig. \ref{fig:g05-1} shows six snapshots in time of the semimayor axis eccentricity plane of the 
evolution of simulation S3, the one we consider is the most representative. At the beginning of the simulation both 
embryos and planetesimals are quickly exited. For embryos this
excitation is due to their own mutual gravitational perturbations, but, as planetesimals are not self-interacting bodies,
their excitation is due to perturbations from embryos. In time, the eccentricities of embryos and planetesimals increase 
until their orbits intersect each other and accretion collisions occur. Then, embryos grow by accretion of other
embryos and planetesimals and the total number of bodies decrease.
By the end of the simulation just a few planets remain in the study region and the system presents the
innermost planet located at 0.72~AU which has a mass of $0.06M_\oplus$, a planet located in the HZ with $0.03M_\oplus$ 
and the most massive planet placed at 2.87~AU with $0.48M_\oplus$. The rest of the simulations present similar results.
 
\begin{figure}[t]
 \centering
 \includegraphics[angle=270, width= 0.45\textwidth]{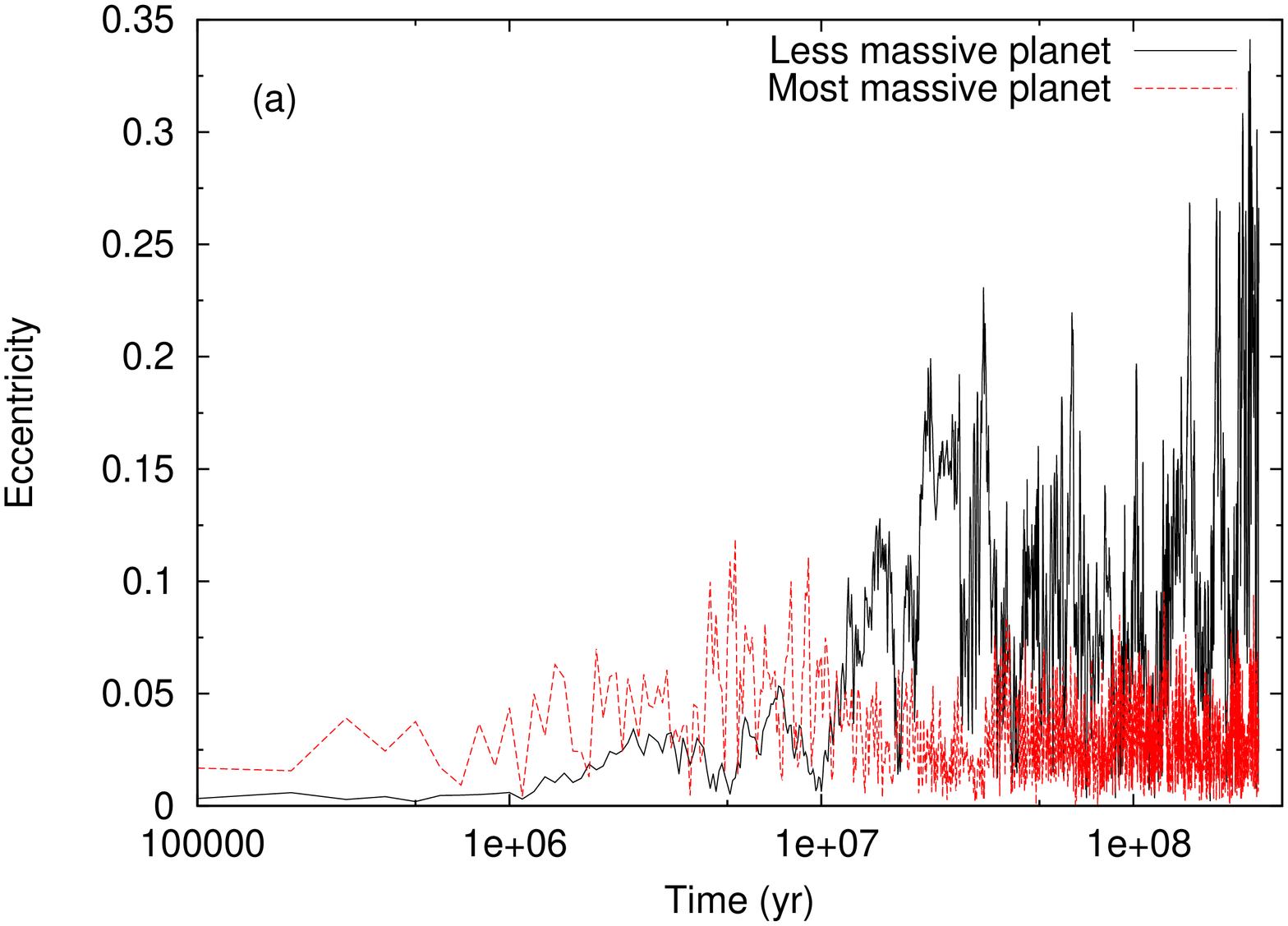}
 \includegraphics[angle=270, width= 0.45\textwidth]{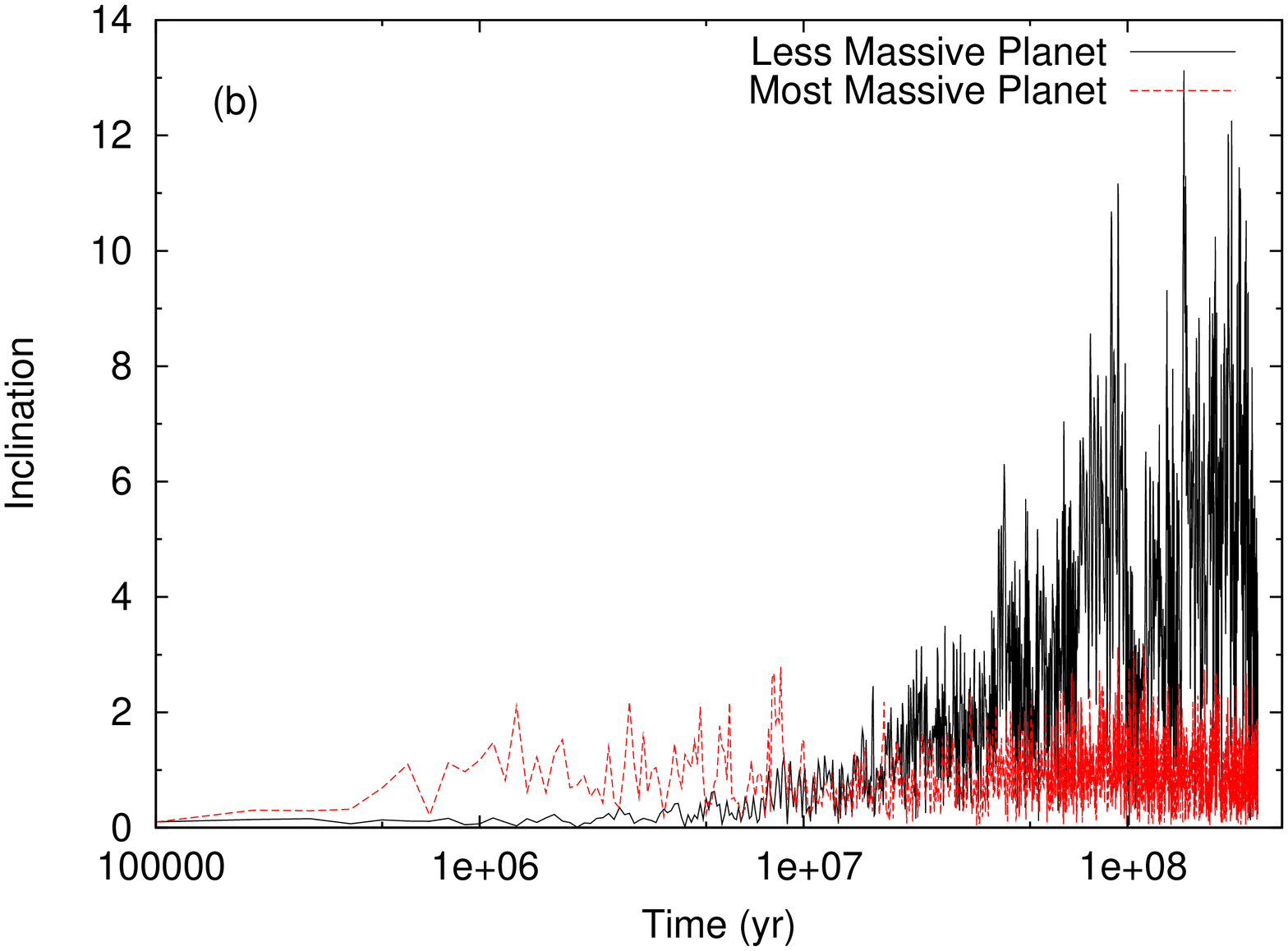}
 \caption{Evolution in time of the eccentricities (a) and inclinations (b) for the less massive planet (black curve) 
and for the most massive planet (red dashed curve) of S3 simulation for $\gamma = 0.5$. The dynamical friction 
phenomena is evident for the most massive planet which does not exceed values of eccentricity 
and inclination of 0.12 and 3.15$^\circ$, respectively. In contrast, the less massive planet reaches maximum values of 
eccentricity and inclination of 0.35 and 13$^\circ$ respectively. Color figure is only available in the electronic version.} 
 \label{fig:g05-2}
\end{figure}

Figure \ref{fig:g05-1} also reveals the importance of the dynamical friction from the beginning of the simulation.
This dissipative force damps the eccentricities and inclinations of the large planetary embryos embedded in a sea of
planetesimals. Particularly, Fig. \ref{fig:g05-2} shows the evolution in time of the eccentricities and inclinations 
of the most and less massive planet of S3 simulation. The less massive planet reaches maximum values of eccentricity and
inclination of 0.35 and 13$^\circ$ respectively, while the most massive planet does not exceed values of eccentricity 
and inclination of 0.12 and 3.15$^\circ$, respectively. Therefore, it is clear that planetesimals are fundamental in order
to describe this phenomena. The three different simulations show similar results concerning the dynamical friction
effects.

After $250~$Myr of evolution many embryos and planetesimals were removed from the disk. The percentage of planetary embryos
and planetesimals that still remain in the study region, between 0.5~AU and 5~AU, is $29\%$ and $46\%$, respectively. These
values represent $1.52M_{\oplus}$ and $0.59M_{\oplus}$, respectively. However,
the last amount of mass in planetesimals only represents $18\%$ of the total initial mass in this 
region. Thus, we can assume this remaining mass in planetesimals will not modify significantly the final planetary system. In 
addition, the remaining mass in planetesimals in the disk at 200~Myr is $0.67M_\oplus$ which represents $20\%$ of the
initial mass. Hence, there are no significant differences after 50~Myr more of evolution.  

Although the orbits of the 2nd. and the 3rd. planets in S3 cross each other, so they could collide and form a 
single planet in the HZ if we extended the simulations, the final mass of it would not exceed $0.07M_\oplus$. In section 5, we will
discuss about the requirements for a planet to be potencially habitable taking on account its final mass.

The most important mass removal mechanism both for embryos and planetesimals is the mass accretion. With this
density profile no embryo collides with the central star and none of them is ejected from the system. Regarding 
planetesimals, only $2.2\%$ of them collide with the central star and a $0.1\%$ ( only one planetesimal) is ejected from the 
disk. These results are consistent with this scenario as it is the least massive profile in the inner zone of the disk.

\begin{figure}[t]
 \centering
 \includegraphics[angle=0, width= 0.45\textwidth]{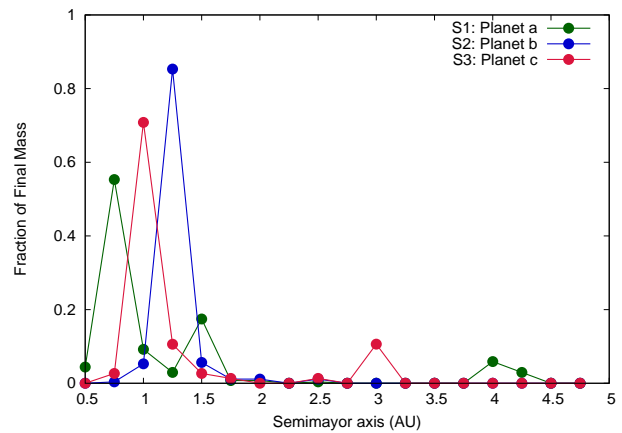}
 \caption{Feeding zones of the planets that remain in the HZ of S1, S2 and S3 in disks with $\gamma = 0.5$. 
          The \emph{y} axis represents the fraction of each planet's final mass after 250~Myr. As can be seen, most of the mass 
          acreted by these planets comes from the inner zone of the disk. Color figure is 
          only available in the electronic version.}
 \label{fig:g05-4}
\end{figure}

All the simulations developed form planets in the HZ and one of the most important results to note is that none of
them come from the outer zone of the disk as we can see in table \ref{tab:2}. Since the surface density profile with 
$\gamma = 0.5$ is the less massive
one in the inner zone of the disk, the gravitational interactions between bodies in this region are weak and
there is not a substantial mixing of solid material. Thus, embryos evolve very close to their initial positions and 
they do not migrate from the outer zone to the inner zone. Because of this, they acrete most of the embryos and planetesimals from the inner zone than from 
the outer zone, beyond the snow line. This can be seen in Fig. \ref{fig:g05-4} where the feeding zones of the planets that remain
in the HZ of S1, S2 and S3 simulations are represented. The $91,16\%$ of the mass of planet \emph{a} was originally situated before 
the snow line. Almost the total mass of planet \emph{b} ($99.9\%$) comes from the inner zone and for planet \emph{c}, the $89.39\%$ 
of the mass comes also from the zone before 2.7~AU. 

Regarding water contents, all planets in the HZ present between $0.06\%$ and $5.37\%$ of water by mass. Particularly, 
the planet in S1 with $0.1M_\oplus$ has $\sim$ 16 Earth's oceans. The planet in S2 with a mass of $0.1M_\oplus$ presents 
$\sim$ 0.2 Earth's oceans and finally the planet in S3 with $0.03M_\oplus$ has $\sim$ 6 Earth's oceans.
Table \ref{tab:2} shows general characteristics of the planets in the HZ for the three
simulations S1, S2 and S3.

It is important to highlight that planetesimals play a protagonist role in these simulations as they are the primary 
responsible of the water content in the resulting planets. 
In fact, almost the total water content in the HZ planet for S1 is provided by planetesimals. However, planetesimals only 
provide the $37.86\%$ of the total mass while the rest of the accreted mass is due to other embryos accretion. The same applies 
to S2 and S3 simulations.

\begin{table*}[t!]
\caption{General characteristics of the planets in the HZ for simulations S1, S2 and S3 for $\gamma =0.5$. $a_{\textrm{i}}$
and $a_{\textrm{f}}$ are the initial and the final semi-mayor axis of the resulting planet in AU, respectively, $M$ the 
final mass in $M_\oplus$, $W$ the percentage of water by mass after $250~$Myr and $T_{\textrm{LGI}}$ the timescale in Myr of
the last giant impact. The planet in the HZ of S3 have not yet present a giant impact; it has only acreted planetesimals. 
As we mentioned in section 4.1, this planet could collide with the 3rd. planet of the system if we extended the simulations beyond 250~Myr. }
\begin{center}
\begin{tabular}{|c|c|c|c|c|c|c|c|}

\hline 
\hline

Simulation  &$a_{\textrm{i}}$(AU) & $a_{\textrm{f}}$(AU) & $M$($M_\oplus$)       & $W$($\%$) & $T_{\textrm{LGI}}$ (Myr) \\
\hline
\hline
S1          &  0.95              & 0.94             & 0.1                & 4.44                    &  146 \\
\hline
S2          &  1.28              & 1.36             & 0.1                & 0.06                    &  49\\
\hline
S3          &  1.20              & 1.15             & 0.03                & 5.37                    &  ---\\
\hline
\hline

\end{tabular}
\end{center}
\label{tab:2}
\end{table*}

\subsection{Simulations with $\gamma = 1$}

We perform three simulations for 300~Myr. At that time the planetesimals that still remain in the disk represent 
a small fraction of the original number, thus they will not modify significantly the final results.
The final planetary systems present 6 or 7 planets with a total mass ranging between $4M_\oplus$ and $4.45M_\oplus$, 
and the masses of individual 
planets is between $0.04M_\oplus$ and $1.9M_\oplus$. Here, all simulations present a planet in the HZ with a range
of masses between $0.18M_\oplus$ and $0.52M_\oplus$ and with water contents from $2.54\%$ to $9\%$ respect the total 
mass which represents from 34 to 167 Earth's oceans.

\begin{figure*}[ht]
 \centering
 \includegraphics[angle=0, width= 0.45\textwidth]{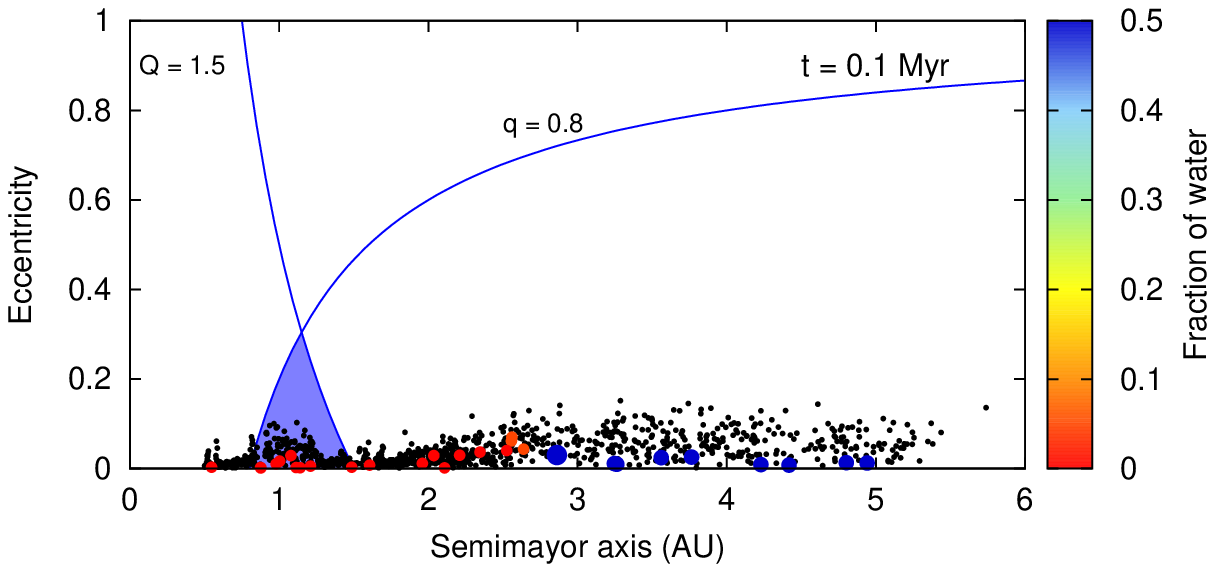}
 \includegraphics[angle=0, width= 0.45\textwidth]{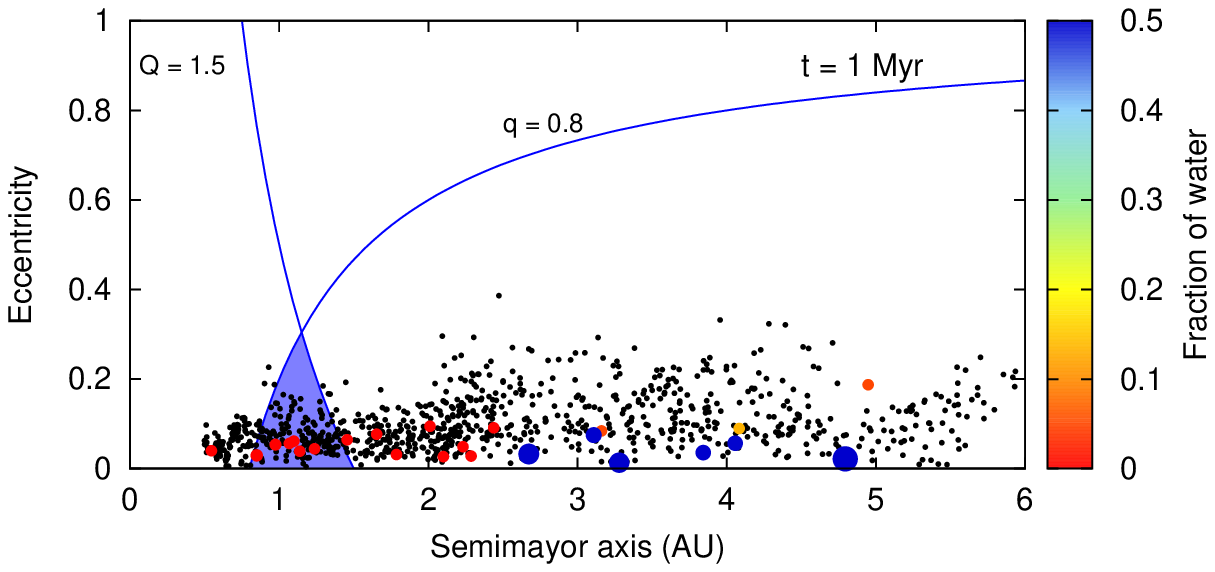}\\
 \includegraphics[angle=0, width= 0.45\textwidth]{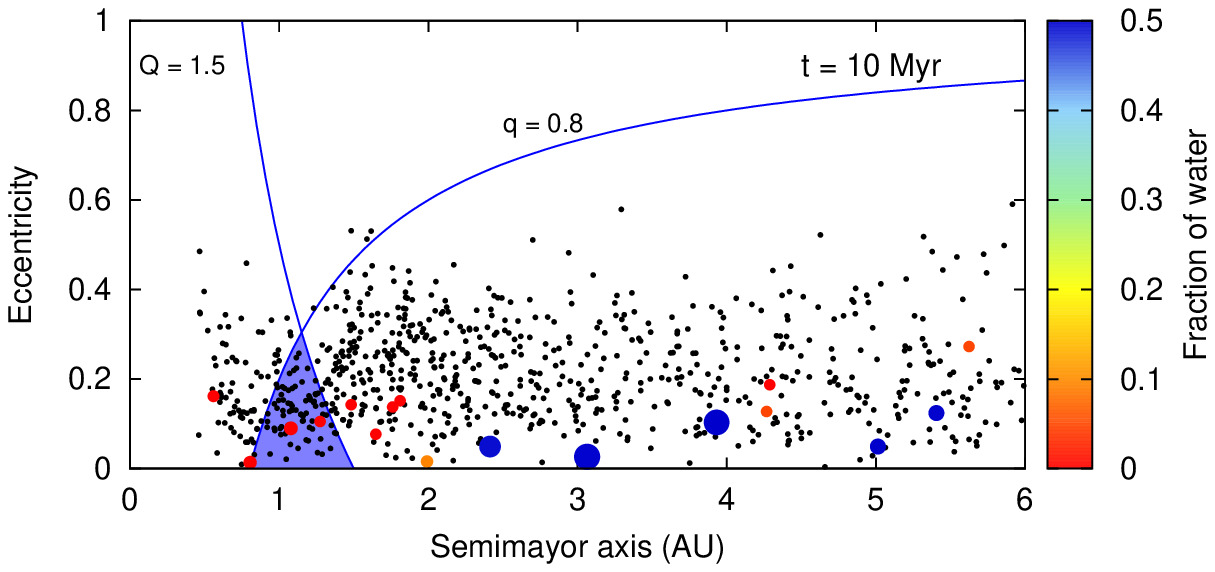}
 \includegraphics[angle=0, width= 0.45\textwidth]{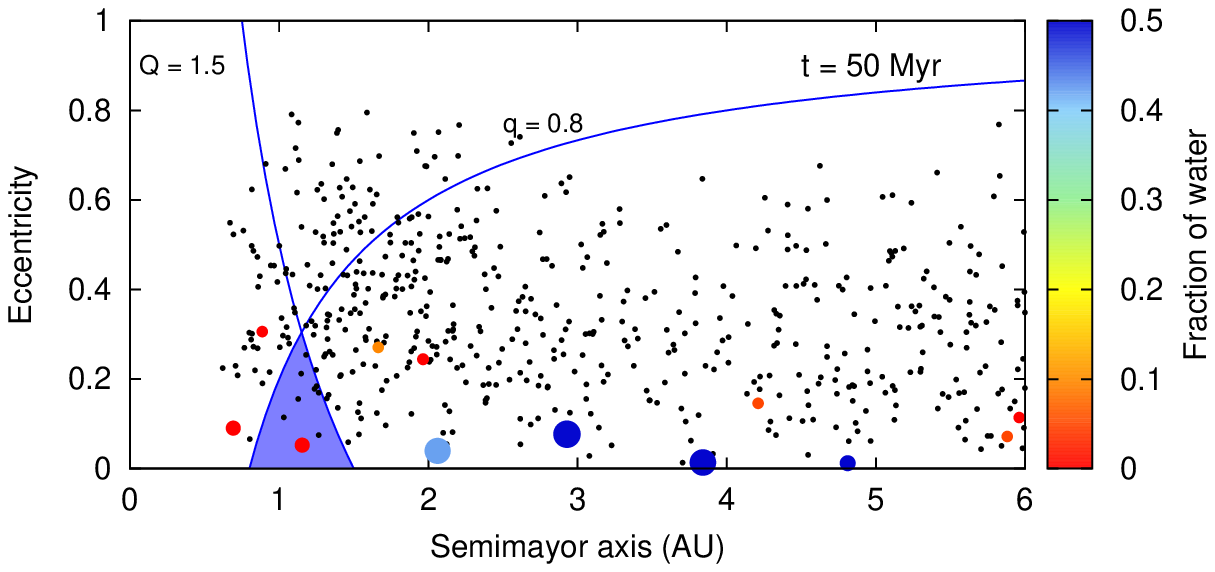}\\
 \includegraphics[angle=0, width= 0.45\textwidth]{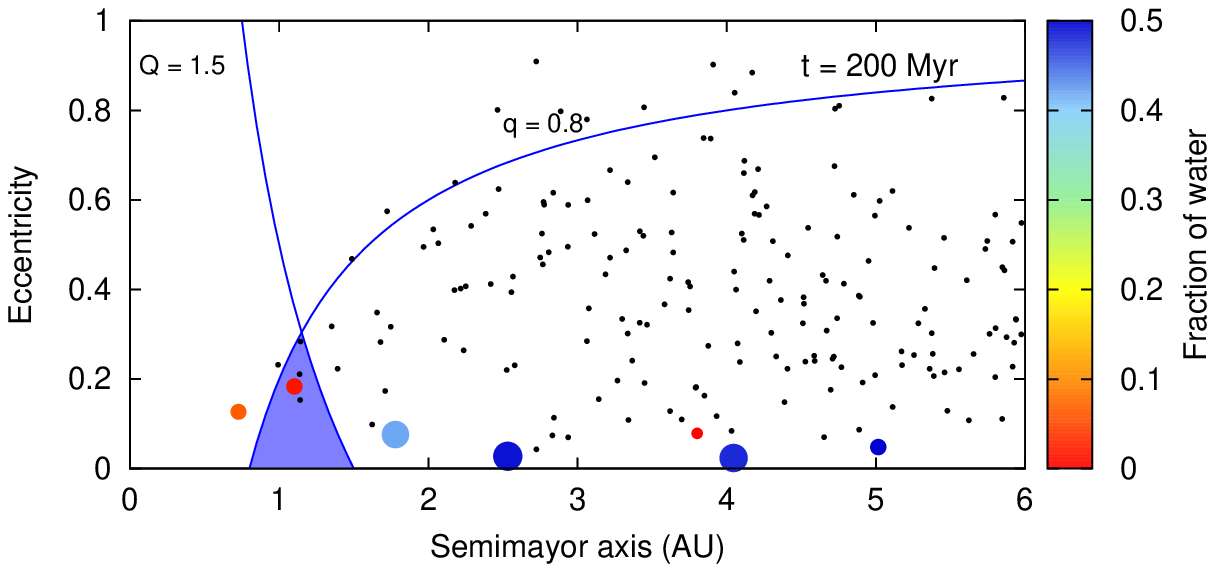}
 \includegraphics[angle=0, width= 0.45\textwidth]{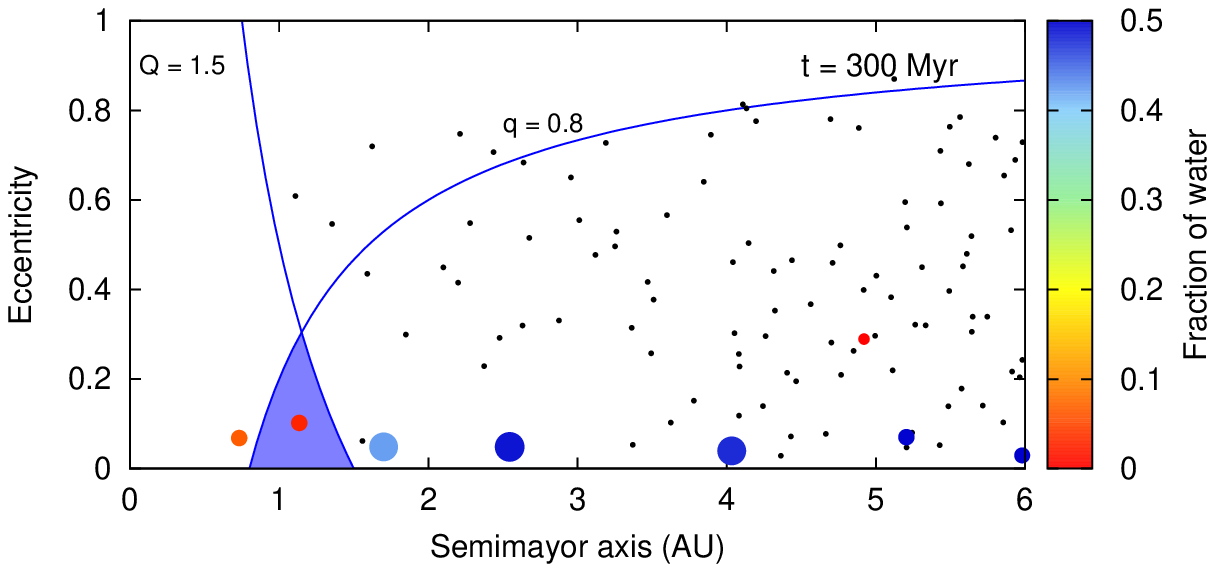}
 \caption{Evolution in time of the S3 simulation for $\gamma = 1$.The light-blue shaded
area represents the HZ and the curves with $q = 0.8$~AU and $Q = 1.5~$AU represents curves of constant perihelion and
aphelion, respectively. Planetary embryos are plotted as colored circles and planetesimals are plotted with black dots. 
The color scale represents the fraction of water of the embryos relative to their masses. For this profile
there is only one embryo in the HZ, located at 1.13~AU with a mass of $0.37M_\oplus$ and a water 
content of $2.54\%$ by mass which represents 34 Earth oceans. Color figure is only available in the electronic version.} 
 \label{fig:g1-1}
\end{figure*}

Figure \ref{fig:g1-1} illustrates six snapshots in time of the semimayor axis eccentricity plane of the 
evolution of S3 simulation, the one we consider is the most representative for this density profile. The accretion 
process is similar to that described for the $\gamma = 0.5$ profile. At the end of the simulation the most important 
characteristics 
of the system can be described as follows. The system presents the innermost planet located at 0.73~AU which has a mass 
of $0.36M_\oplus$, a planet located in the HZ with $0.37M_\oplus$ and the most massive planet placed at 2.54~AU with 
$1.26M_\oplus$. The rest of the simulations present similar results.

\begin{figure}[ht]
 \centering
 \includegraphics[angle=270, width= 0.45\textwidth]{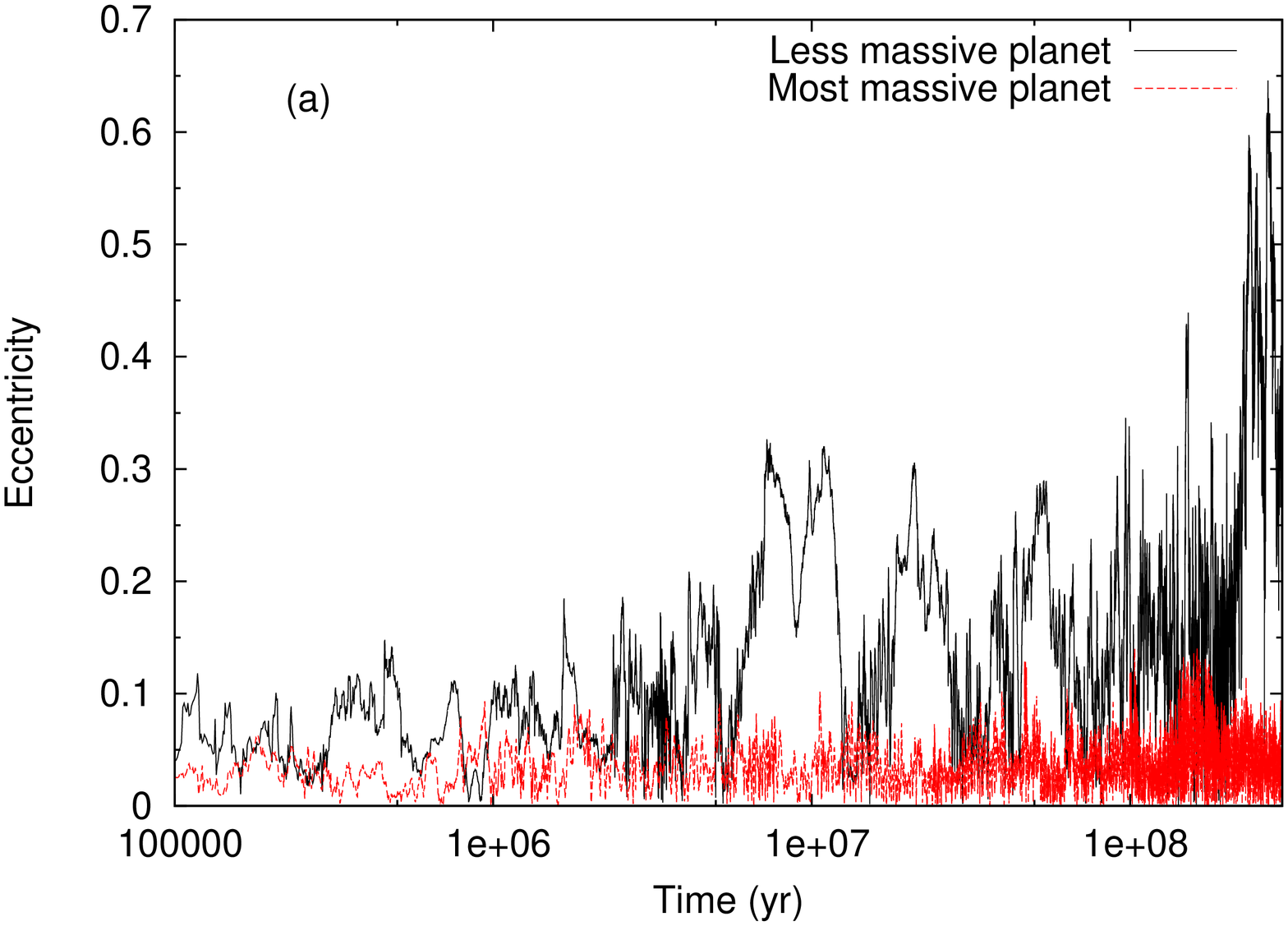}
 \includegraphics[angle=270, width= 0.45\textwidth]{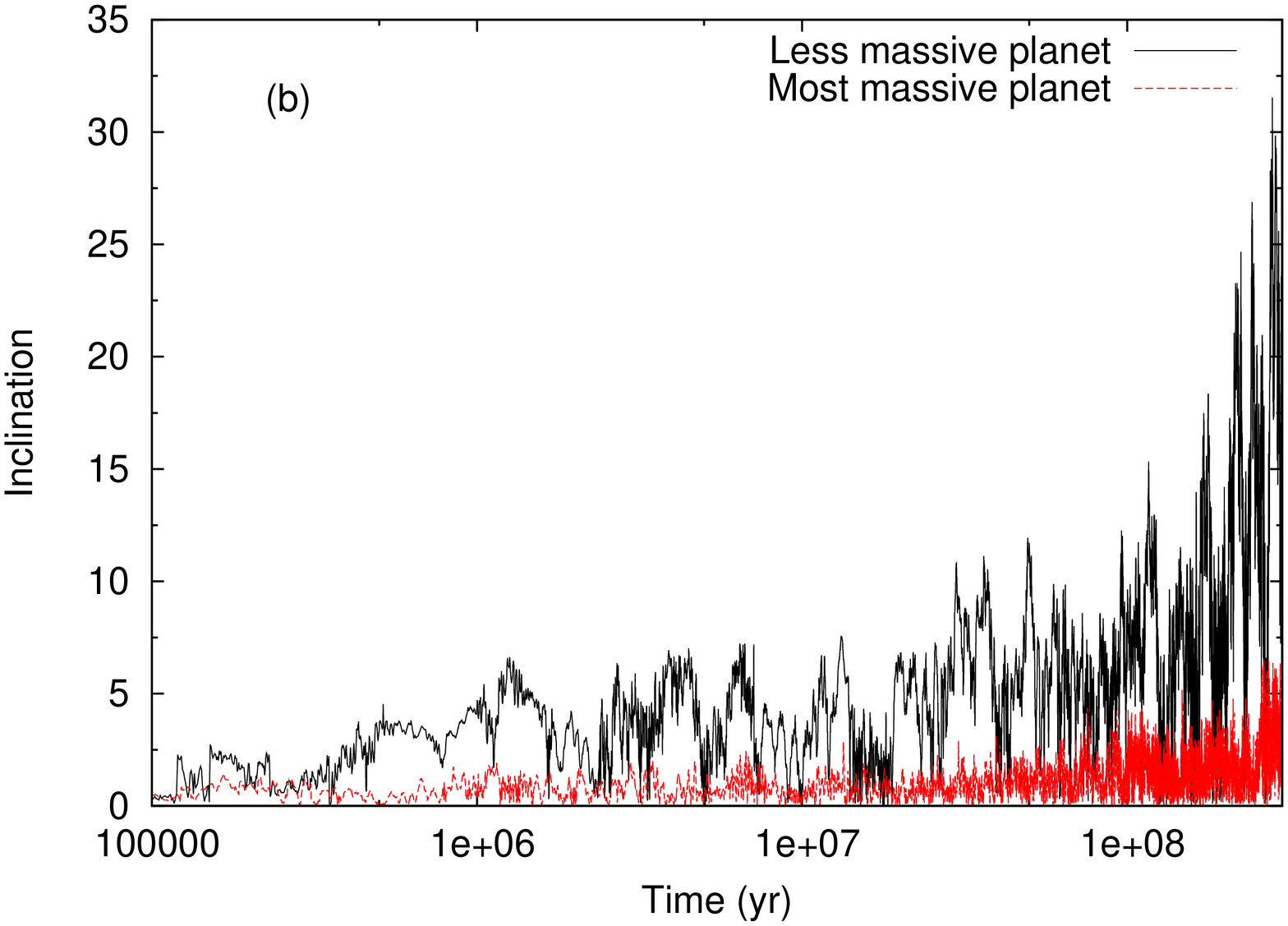}
 \caption{Evolution in time of the eccentricities (a) and inclinations (b) for the less massive planet (black curve) 
and for the most massive planet (red dashed curve) of S3 simulation for $\gamma = 1$. The dynamical friction 
phenomena is evident for the most massive planet which presents damped eccentricity and inclination. The scales of
eccentricity and inclination are higher than the ones in the first profile with $\gamma = 0.5$. Color figure is 
only available in the electronic version. } 
 \label{fig:g1-2}
\end{figure}

The dynamical friction effects are also relevant in this density profile from the beginning of the simulation and 
for the most massive bodies. Inclinations and eccentricities of the most massive bodies are damped by this dissipative force.
Figure \ref{fig:g1-2} shows for S3 simulation, the evolution in time of the eccentricities and inclinations of the
most and less massive planet. The less massive planet reaches maximum values of eccentricity and
inclination of 0.64 and 31.53$^\circ$ respectively, while the most massive planet does not exceed values of eccentricity 
and inclination of 0.14 and 6.48$^\circ$, respectively. The difference with the same results for $\gamma = 0.5$ is that
the scales of eccentricity and inclination are higher due to the fact that this profile is more massive than the first one.
All simulations for this profile show similar results concerning this phenomena.

With respect to the final number of resulting embryos and planetesimals, we obtain that 
$\sim 6.2\%$ of planetesimals still survive in the study region after 300~Myr, while regarding embryos, a $20\%$ is still 
in the disk. These values represent $0.28M_\oplus$ and $4.46M_\oplus$ in planetesimals and embryos, respectively.
Although there is still solid mass to continue the accretion process, this remaining mass in planetesimals only 
represents a $3.53\%$ of the initial mass in the study region. Therefore, it will not provide significant differences in
the final planetary system.
For this density profile the most important mass removal mechanism remains the mass accretion since no embryo collides 
with the central star and none of them is ejected from the system. As to the mass in planetesimals, a $16.4\%$ 
collides with the central star and a $11.5\%$ is ejected from the disk. In this profile, contrast to the previous one,
the dynamical excitation is more evident since is more massive in the inner zone of the disk, providing a 
higher percentage of planetesimals ejections from the system and collisions with the central star.

\begin{figure}[ht]
 \centering
 \includegraphics[angle=0, width= 0.45\textwidth]{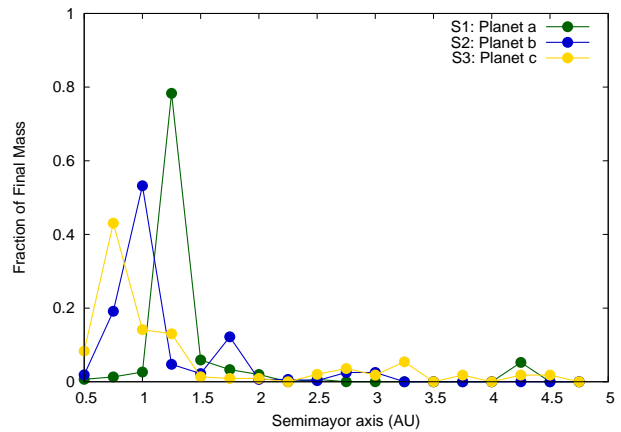}
 \caption{Feeding zones of the planets that remain in the HZ of S1, S2 and S3 in disks with $\gamma = 1$. 
          The \emph{y} axis represents the fraction of each planet's final mass after 300~Myr. Here again, as can be seen, most of 
          the mass acreted by these planets comes from the inner zone of the disk. Color figure is 
          only available in the electronic version.}
 \label{fig:g1-4}
\end{figure}

As we have already said, one of the points of interest in this work is to study the planets that remain in the HZ. This profile 
also forms planets in the HZ and again, none of them come from the outer zone of the disk (see table \ref{tab:3}). 
Despite this density profile with value of $\gamma = 1$ presents more mass in the inner zone, it is still not enough to produce 
strong gravitational interactions between bodies, and therefore the mix of solid material is very low. Due to this, embryos evolve close 
to their initial positions and they do not migrate from the outer to the inner zone.
These planets acrete most of the embryos 
and planetesimals from before the snow line. This can be seen in Fig. \ref{fig:g1-4} where the feeding zones of the planets that remain
in the HZ of S1, S2 and S3 simulations are represented. In this case, the $83.75\%$ of the mass of planet \emph{a} was originally situated before 
2.7~AU. The $94.74\%$ of the total mass of planet \emph{b} comes from the inner zone and for planet \emph{c}, the $94.95\%$ 
of the mass also comes from the inner zone. 

All planets resulting in the HZ for the three simulations present between $2.54\%$ and $9\%$ of water by mass.
In particular, the planet in S1 with $0.52M_\oplus$ has $\sim$ 167 Earth's oceans. The planet in S2 with a mass of 
$0.18M_\oplus$ presents $\sim$ 17 Earth's oceans and the planet in S3 with $0.37M_\oplus$ has $\sim$ 34 Earth's oceans.
 Finally, Table \ref{tab:3} shows general characteristics of the planets in the HZ for the three
simulations S1, S2 and S3.

\begin{table*}[t!]
\caption{General characteristics of the planets in the HZ for simulations S1, S2 and S3 for $\gamma =1$. $a_{\textrm{i}}$
and $a_{\textrm{f}}$ are the initial and the final semimayor axis of the resulting planet in AU, respectively, $M$ the 
final mass in $M_\oplus$, $W$ the percentage of water by mass after $300~$Myr and $T_{\textrm{LGI}}$ the timescale in Myr of
the last giant impact.}
\begin{center}
\begin{tabular}{|c|c|c|c|c|c|c|c|}

\hline
\hline

Simulation  &$a_{\textrm{i}}$(AU) & $a_{\textrm{f}}$(AU) & $M$($M_\oplus$)       & $W$($\%$) & $T_{\textrm{LGI}}$ (Myr) \\
\hline
\hline
S1          &  0.84              & 0.83             & 0.52                & 9.00                    & 110 \\
\hline
S2          &  1.41              & 1.15             & 0.18                & 2.66                    & 2 \\
\hline
S3          &  0.99              & 1.02             & 0.37                & 2.54                    & 78 \\
\hline
\hline

\end{tabular}
\end{center}
\label{tab:3}
\end{table*}
In general, all these planetary systems also show that those responsible for the final water content are the 
planetesimals which provide almost the total water content in the HZ planets. Nevertheless, they only provide 
between $30\%$ to $50\%$ of the final mass of the planet.

\subsection{Simulations with $\gamma = 1.5$}

Finally we describe the planetary systems with $\gamma = 1.5$. As this is the most massive density profile 
in the inner zone of the disk is expected to be the most distinctive one, since the mass of the inner 
zone favors the formation of planets in the HZ. 

We perform three different simulations
and integrate them for 200~Myr. At the end of the simulations, there is almost no extra mass to continue the accretion
process and thus, this result suggest that these systems have reached a dynamical stability. 
The most relevant characteristics of the three simulations we perform can be listed as follows. At the end of the 
simulations, the study region contains between 4 and 7 planets with a total mass from $6.97M_\oplus$
to $8.81M_\oplus$. The mass of the final planets ranges between $0.06M_\oplus$ to $3.08M_\oplus$ and the resulting planets
in the HZ present masses from $0.66M_\oplus$ to $2.21M_\oplus$. All simulations form a planet in the HZ, particularly 
S2 simulation forms 2 planets. Regarding water contents in the HZ, we find that planets present ranges from $4.51\%$
to $39.48\%$ by mass which represent from 192 to 2326 Earth's  oceans. 
 
\begin{figure*}[ht]
 \centering
 \includegraphics[angle=0, width= 0.45\textwidth]{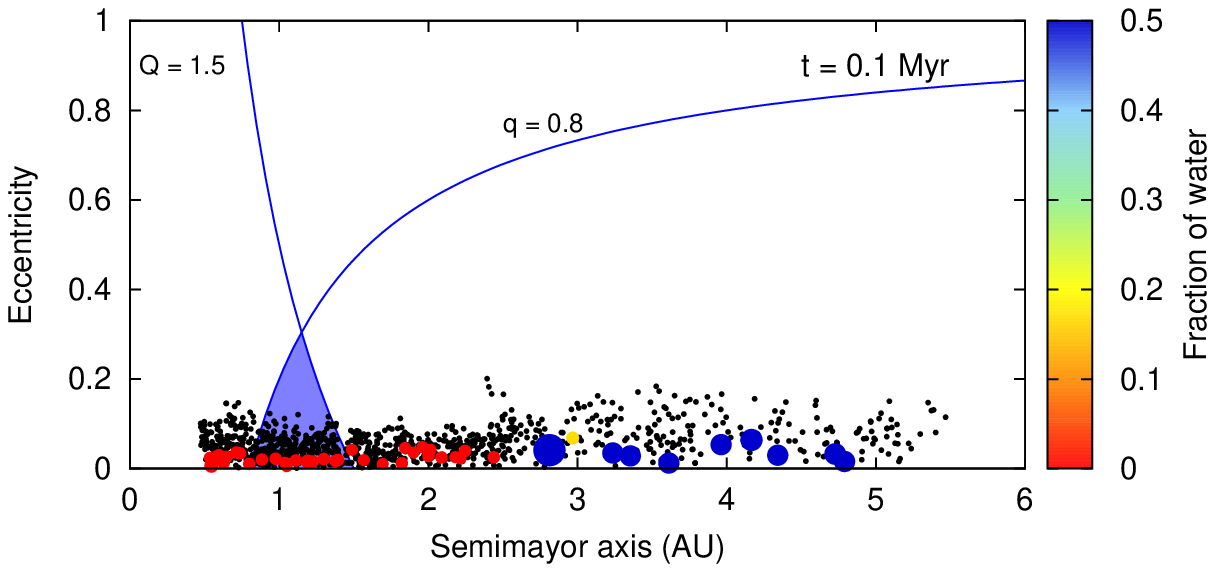}
 \includegraphics[angle=0, width= 0.45\textwidth]{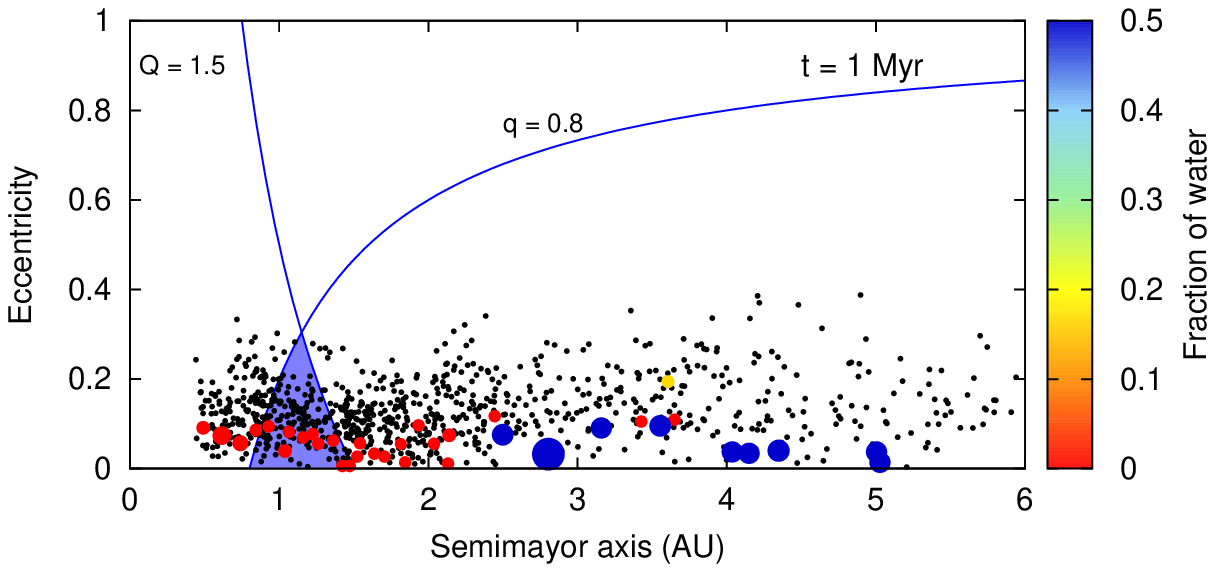}\\
 \includegraphics[angle=0, width= 0.45\textwidth]{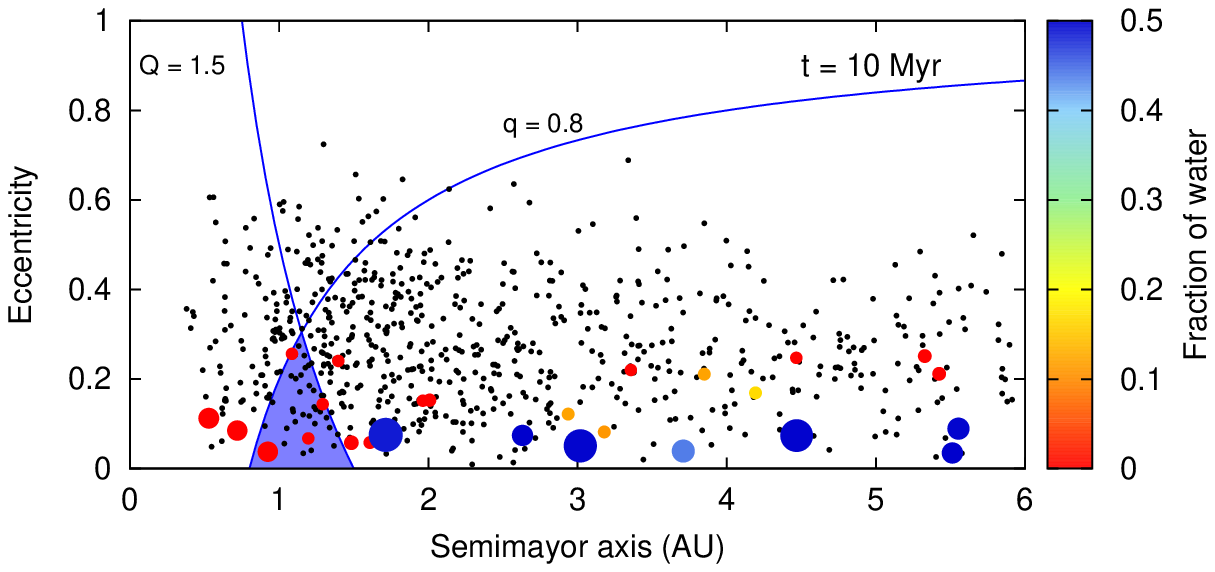}
 \includegraphics[angle=0, width= 0.45\textwidth]{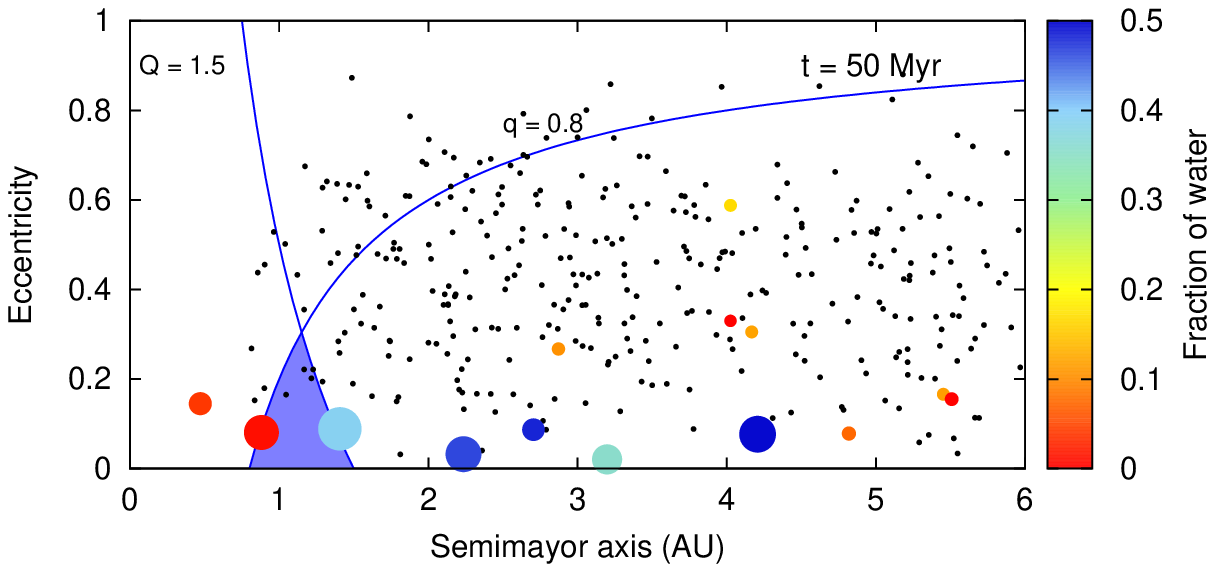}\\
 \includegraphics[angle=0, width= 0.45\textwidth]{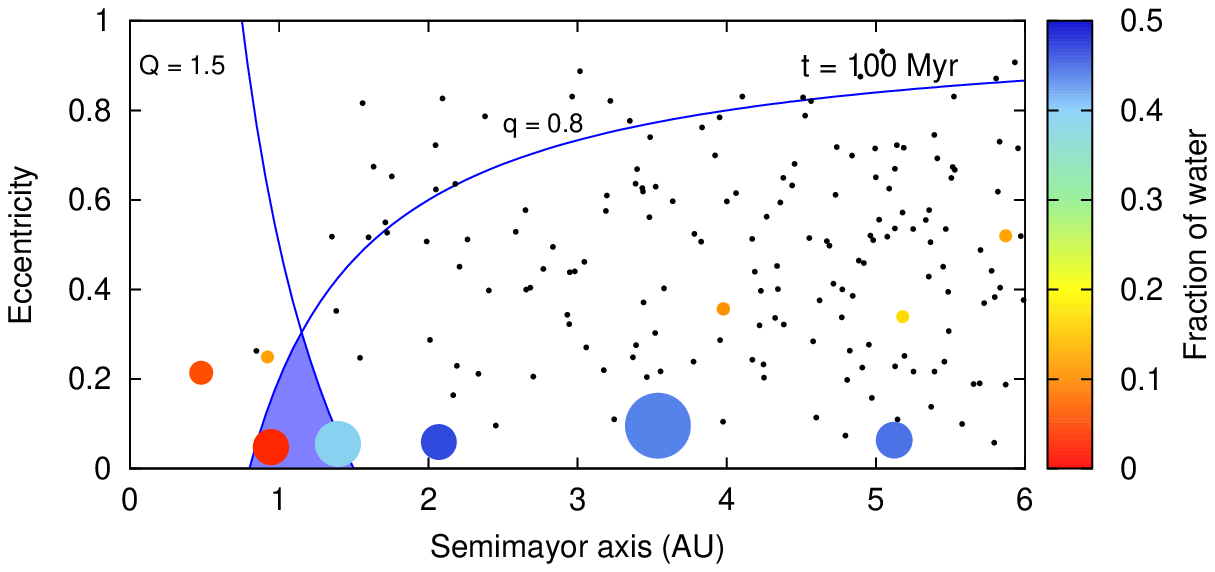}
 \includegraphics[angle=0, width= 0.45\textwidth]{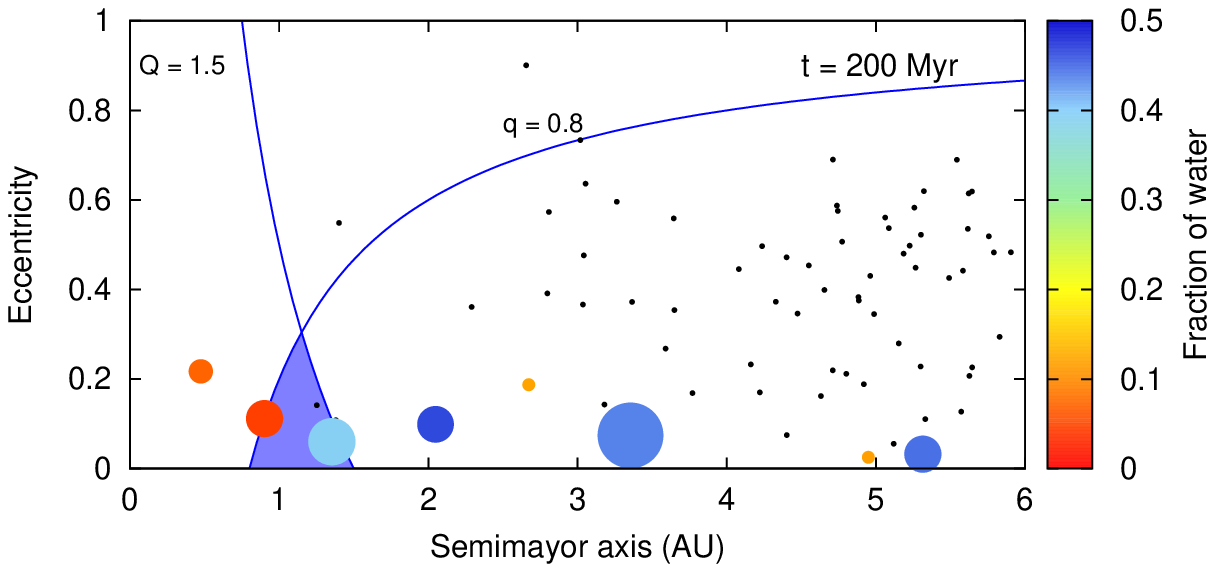}\\
 \caption{Evolution in time of the S2 simulation for $\gamma = 1.5$. The light-blue shaded
area represents the HZ and the curves with $q = 0.8$~AU and $Q = 1.5~$AU represents curves of constant perihelion and
aphelion, respectively. Planetary embryos are plotted as colored circles and planetesimals are plotted with black dots. 
The color scale represents the fraction of water of the embryos respect to their total masses. In 
this case there are two planets in the HZ with masses of $1.19M_\oplus$ and $1.65M_\oplus$, respectively. 
They present $4.51\%$ and $39.48\%$ of water content by mass, which represents $192$ and $2326$ Earth's oceans, respectively. 
Color figure is only available in the electronic version.} 
 \label{fig:g15-1}
\end{figure*}

S2 simulation is the most interesting one for this density profile because presents 2 planets in the HZ. This is why
we chose this simulation for a more detailed description. Figure \ref{fig:g15-1} shows six snapshots in time of the 
semimayor axis eccentricity plane of the evolution of S2. In this case, the dynamical excitation of eccentricities and 
inclinations of both planetary embryos and planetesimals increases faster than in the other two described profiles. This 
promotes the mix of solid material between the inner and outer region of the snow line. After 200~Myr of
evolution, the planetary system shows the innermost planet located at 0.47~AU with a mass of $0.60M_\oplus$, 2 planets
in the HZ with $1.19M_\oplus$ and $1.65M_\oplus$ and the most massive planet placed at 3.35~AU which has a mass
of $2.15M_\oplus$. 

\begin{figure}[ht]
 \centering
 \includegraphics[angle=270, width= 0.45\textwidth]{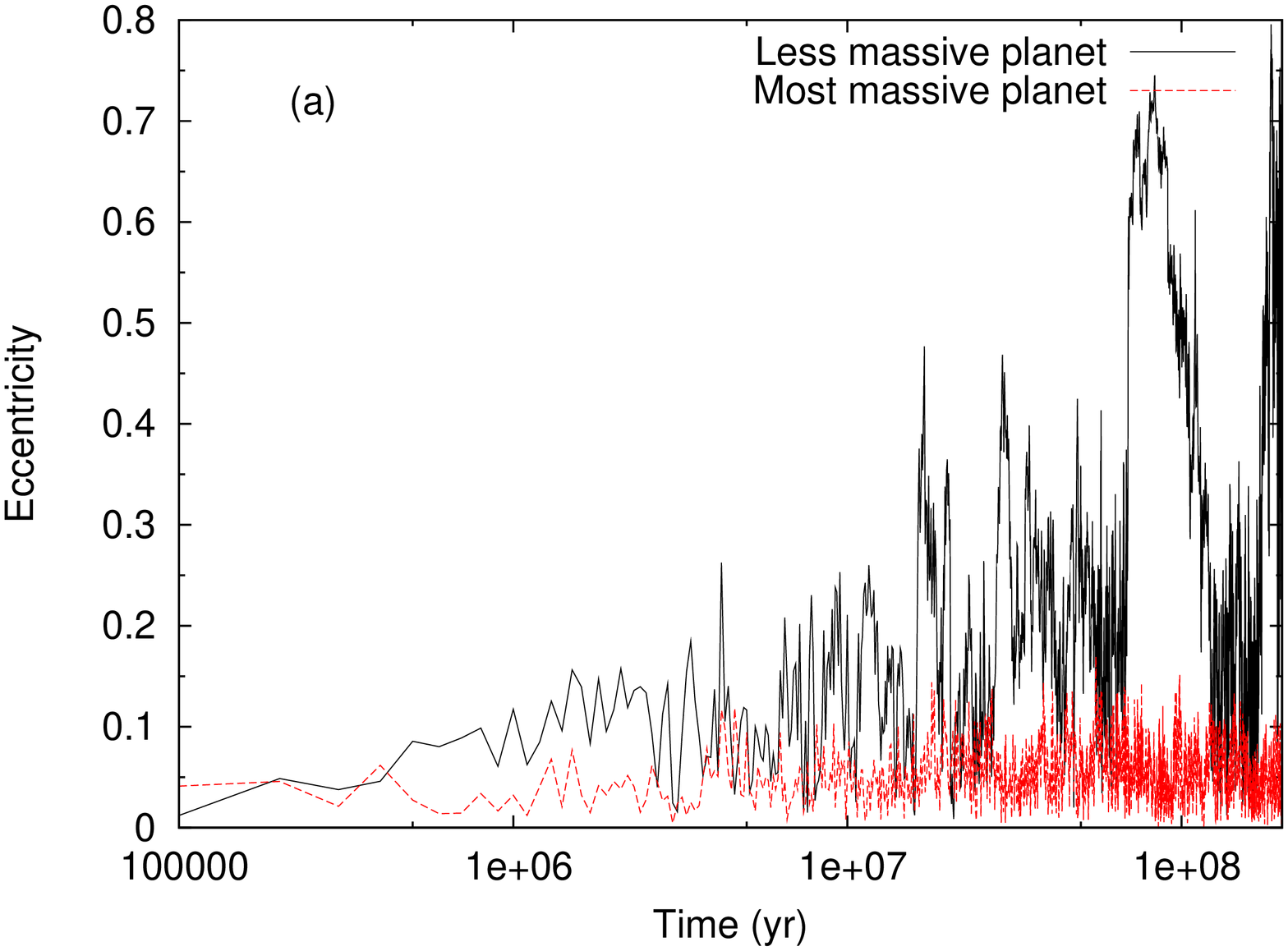}
 \includegraphics[angle=270, width= 0.45\textwidth]{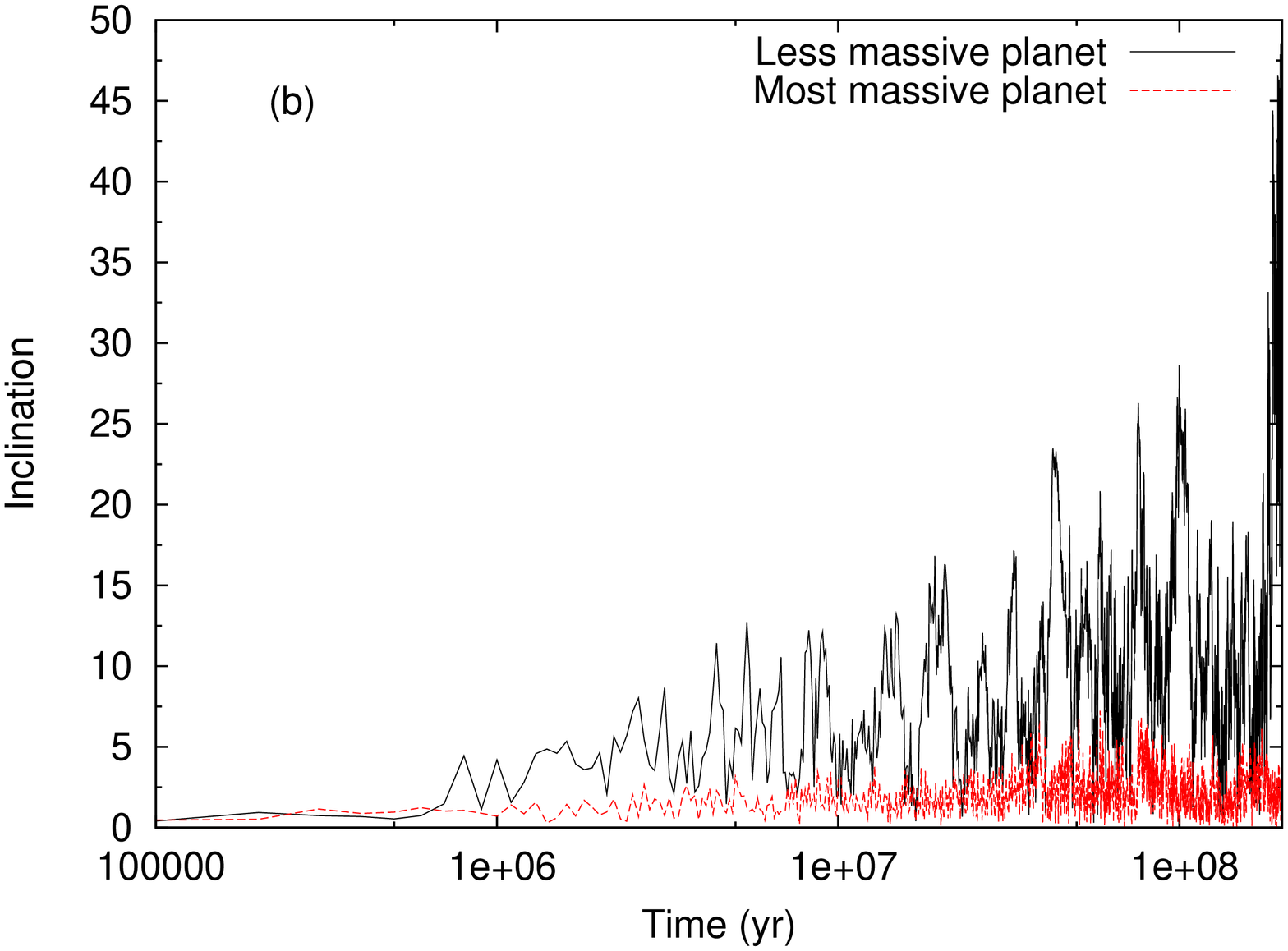}
 \caption{Evolution in time of the eccentricities (a) and inclinations (b) for the less massive planet (black curve) 
and for the most massive planet (red dashed curve) of S2 simulation for $\gamma = 1.5$. The dynamical friction 
phenomena is evident for the most massive planet which presents damped eccentricity and inclination. The scales of
eccentricity and inclination are higher than those for $\gamma = 0.5$ and 1. Color figure is 
only available in the electronic version.} 
 \label{fig:g15-2}
\end{figure}

This profile presents the most massive planets and this further evidences the effects of dynamical friction. Indeed, 
the less massive planet for S2 simulation reaches values of eccentricity and inclination of 0.79 and $48.56^{\circ}$, 
respectively, while the most massive planet present maximum values of eccentricity and inclination of 0.17 and 
$7.23^{\circ}$, respectively. This is illustrated in Fig. \ref{fig:g15-2}. Therefore, as shown, it is still a tendency 
that the effects of dynamical friction prevail over larger bodies. The main difference with the same results for the profiles
with $\gamma = 0.5$ and $\gamma = 1$ is that the scales of eccentricities and inclinations are higher. 
All simulations for this surface density profile present similar results concerning this phenomena.

S2 simulation ends with a $4.1\%$ of survival planetesimals after 200~Myr between 0.5~AU and 5~AU, while regarding embryos, 
a $15.5\%$ is still in the disk. These values represent $0.43M_\oplus$ and $6.97M_\oplus$ in planetesimals and embryos, 
respectively.
The remaining mass in planetesimals is not enough to modify significantly the final planetary system.
This follows that the formation scales we use with this surface 
density profile are suitable. The most important mass removal mechanism is, again, the mass accretion. No 
embryo collides with the central star and none of them is ejected from the system. However, a $23.4\%$ of the 
planetesimals collides with the central star and a $11.9\%$ is ejected from the disk. This shows that the gravitational 
interactions between bodies are stronger than in the other profiles because the mass in the inner zone is greater
than that associated to profiles with $\gamma = 0.5$ and $1$.

\begin{figure}[ht]
 \centering
 \includegraphics[angle=0, width= 0.45\textwidth]{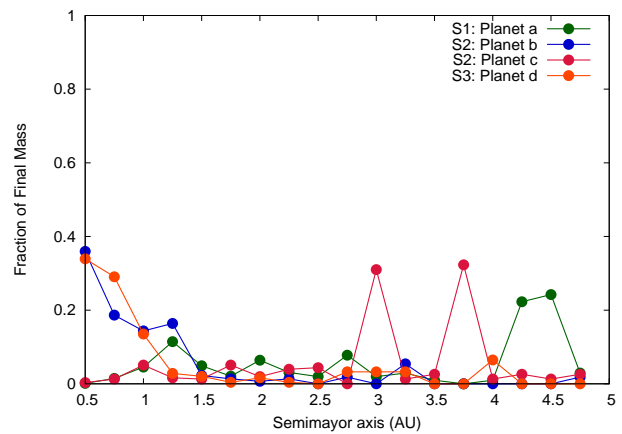}
 \caption{Feeding zones of the planets that remain in the HZ of S1, S2 and S3 in disks with $\gamma = 1.5$. 
          The \emph{y} axis represents the fraction of each planet's final mass after 200~Myr. In this case, most of 
          the mass acreted by planets \emph{b} and \emph{d} comes from the inner zone of the disk. But for planets \emph{a} and \emph{c} is the
          other way arround. Color figure is only available in the electronic version.}
 \label{fig:g15-4}
\end{figure}

All simulations, as we said, present a planet in the HZ, particularly one of them presents two. But this profile presents
a distinctive and interesting characteristic from the others. Two of the four planets that remain in the HZ come from the
outer zone of the disk, this is, they come from beyond the snow line. This situation is due to the solid material mix 
which is stronger than in the other two profiles. As this profile is the more massive one in the inner zone, strong 
gravitational interactions are favored between bodies. This migration of the planets in the HZ for S2 simulation can be
seen in Table \ref{tab:4}. This situation is also 
represented in \ref{fig:g15-4} with the feeding zones 
of the planets that remain in the HZ of S1, S2 and S3. Here, only the $35.98\%$ of the mass of planet \emph{a} was originally situated before 
2.7~AU, the rest of the mass comes from beyond the snow line. Something similar happens with planet \emph{c} where only the $25\%$ of the mass 
comes from the inner zone. Then, for planets \emph{b} and \emph{d} the $90.98\%$ and the $83.79\%$ of the total mass, respectively, comes 
from inside the snow line.

Finally, this seem to be the most peculiar profile because presents \emph{water worlds}, which are planets with high percentages
of water contents by mass. All planets that remain in the 
HZ for the three simulations present between $4.51\%$ and $39.48\%$ of water content respect to the total mass, being 
the embryos which come from beyond the snow line the ones that present large amounts of water.
Particularly, the planet in S1 with $2.21M_\oplus$ has $32.55\%$ of water content which represents $\sim$ 2569 Earth's 
oceans. This planet comes from the outer disk. The planets in S2 simulation with masses of $1.19M_\oplus$ and 
$1.65M_\oplus$ present $4.51\%$ and $39.48\%$ of water content, respectively, which equal to 192 and 2326 Earth's oceans, 
respectively. The first planet comes from the inner zone of the disk and the second one 
comes from the outer zone. In S3 simulation, the planet in the HZ, which comes from the inner zone of the disk, has 
$0.66M_\oplus$ and shows a $8.10\%$ of water by mass, which represents $\sim$ 191 Earth's oceans. 
Lastly, Table \ref{tab:4} presents general 
characteristics of the planets in the HZ for the three simulations S1, S2 and S3.

\begin{table*}[t!]
\caption{General characteristics of the planets in the HZ for simulations S1, S2 and S3 for $\gamma =1.5$. $a_{\textrm{i}}$
and $a_{\textrm{f}}$ are the initial and the final semi-mayor axis of the resulting planet in AU, respectively, $M$ the 
final mass in $M_\oplus$, $W$ the percentage of water by mass after $200~$Myr and $T_{\textrm{LGI}}$ the timescale in Myr of
the last giant impact.}
\begin{center}
\begin{tabular}{|c|c|c|c|c|c|c|c|}

\hline
\hline

Simulation  &$a_{\textrm{i}}$(AU) & $a_{\textrm{f}}$(AU) & $M$($M_\oplus$)       & $W$($\%$) & $T_{\textrm{LGI}}$ (Myr) \\
\hline
\hline
S1          &  4.52              & 1.41             & 2.21                & 32.55                   & 32 \\
\hline
S2          &  1.30              & 0.90             & 1.19                & 4.51                    & 35 \\
            &  3.13              & 1.35             & 1.65                & 39.48                   & 22 \\
\hline
S3          &  0.64              & 0.98             & 0.66                & 8.10                    &  6 \\
\hline
\hline

\end{tabular}
\end{center}
\label{tab:4}
\end{table*}

As for the profiles with $\gamma = 0.5$ and $\gamma = 1$, planets in the HZ that do not come from beyond the snow line
owe their water contents almost entirely to planetesimals, being responsible of the $\sim 50\%$ of their final
masses. However, the water content of the planets in the HZ that come from the outer zone is almost due to their initial
water contents. Thus, for water worlds, planetesimals represent a secondary source of water.


\section{Discussion and Conclusions}

We presented results concerning the formation of planetary systems without gaseous giants
around Sun-like stars. In particular, our study assumed a low-mass protoplanetary disk with
0.03 $M_{\odot}$ and considered a wide range of surface density profiles. The choice of such
conditions was based on results derived by \citet{Miguel2011}, who suggested
that a planetary system composed by only rocky planets is the most common outcome obtained
from a low-mass disk (namely, $\lesssim$ 0.03 $M_{\odot}$) for different surface density
profiles. We used a generic surface density profile characterized by a power law in the inner
disk of the form $r^{-\gamma}$ and an exponential taper at large radii \citep{Lynden-Bell1974,Hartmann1998}. 
To describe a wide diversity of planetary systems, we chose values
for $\gamma$ of 0.5, 1, and 1.5. For each of these surface density profiles, we developed three
N-body simulations of high resolution, which combined the interaction between planetary embryos
and planetesimals. These N-body simulations allowed us to describe the dynamical evolution and
the accretion history of each of the planetary systems of our study. In particular, these
simulations analyzed the delivery of water to the resulting planets, allowing us to determine
the importance of the systems under consideration from an astrobiological point of view.

\begin{figure*}[ht]
 \centering
 \includegraphics[angle=0, width= 0.7\textwidth]{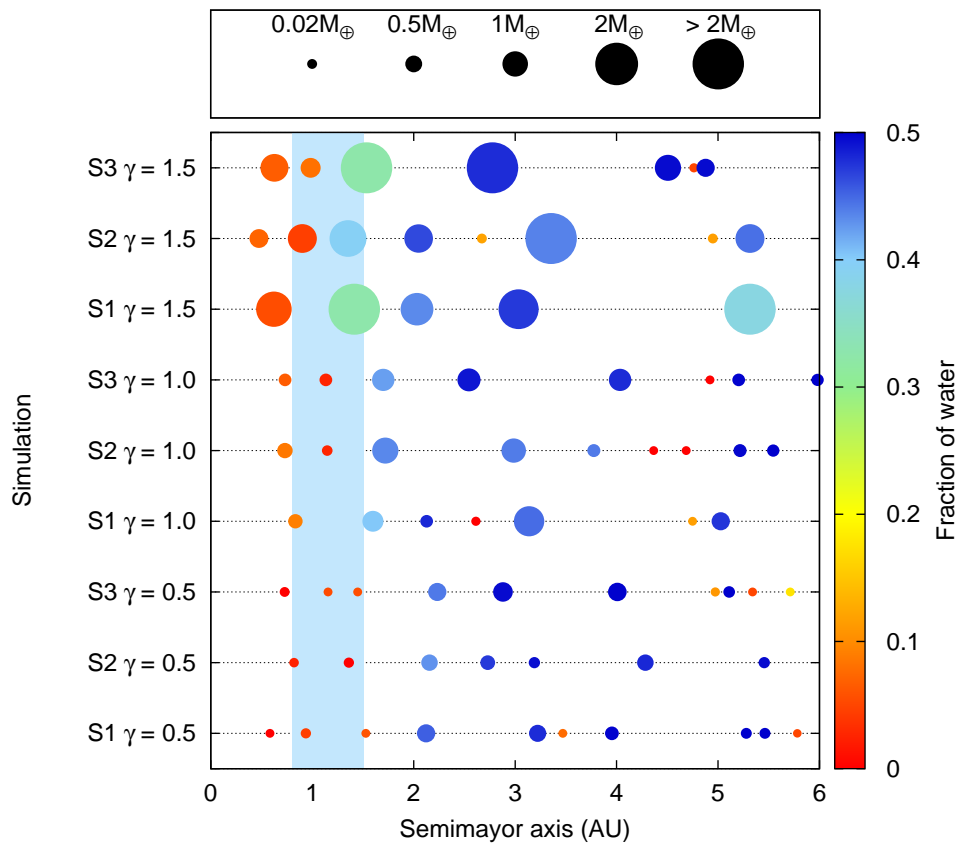}
 \caption{Final configuration of all nine simulations. The color scale represents the water content of each planet and the shaded region 
represents the HZ between 0.8~AU and 1.5~AU. The size of each planet represents its relative physical size for those planets with masses lower 
or equal to $2M_\oplus$.}
 \label{fig:todos}
\end{figure*}

The most interesting planets of our simulations are those formed in the HZ of the system. All
the simulations formed planets in the HZ with different masses and final water contents depending
on the surface density profile. Figure \ref{fig:todos} shows the final configuration of all nine simulations. Here
we can appreciate the diversity of possible planetary systems of terrestrial planets that could form around
solar-type stars without giant gas planets and in low-mass disks.

For $\gamma =$ 0.5, our simulations produced three planets in the
HZ with masses ranging from 0.03 $M_{\oplus}$ to 0.1 $M_{\oplus}$ and water contents between 0.2 and
16 Earth oceans. While these planets are formed in the HZ and present final water contents comparable
and even higher than that of the Earth, their masses do not seem to be large enough to retain a substantial
and long-lived atmosphere neither to sustain plate tectonics. In fact, \citet{Williams1997} proposed that the lower limit
for habitable conditions is $M > 0.23M_\oplus$. Beyond the uncertainties in this value, we believe that such systems 
would not be of particular interest because the planets in the HZ would not be able to reach or to overcome the estimated mass.

For $\gamma =$ 1, three planets formed in the HZ with masses between 0.18 $M_{\oplus}$ and 0.52 $M_{\oplus}$
and water contents ranging from 34 to 167 Earth oceans. Although these final water contents represent
upper limits, we infer that the planets formed in the HZ from this surface density profile are water-rich
bodies. On the other hand, their masses seem to be suitable considering the requirements necessary to retain
a long-lived atmosphere and to maintain plate tectonics \citep{Williams1997}. Thus, we suggest that the planets
produced in the HZ from this surface density profile result to be of astrobiological interest.

For $\gamma =$ 1.5, our simulations formed four planets in the HZ with masses ranging from 0.66 $M_{\oplus}$
to 2.21 $M_\oplus$ and water contents between 192 and 2326 Earth oceans. Taking into account the masses
and the final water contents of such planets, these planetary systems are of special astrobiological interest.
It is worth noting that this surface density profile shows distinctive results since it is the only one of
those analyzed here that forms planets with very high proportion of water relative to the composition
of the entire planet. In fact, two of the planets formed in the HZ are water worlds with masses of 1.65
$M_{\oplus}$ and 2.21 $M_{\oplus}$ and 39.5 \% and 32.6 \% water by mass, respectively. For each of these cases,
an embryo located beyond the snow line at the beginning of the simulation served as the accretion seed for
the potentially habitable planet. It is very important to discuss the degree of interest of such water
worlds from an astrobiological point of view. \citet{Abbot2012} studied the sensitivity of planetary
weathering behavior and habitable zone to surface land fraction. They found that the weathering behavior
is fairly insensitive to land fraction for partially ocean-covered planets, as long as the land fraction
is greater than $\sim$ 0.01. Thus, this study suggests that planets with partial ocean coverage should
have a habitable zone of similar width. On the other hand, \citet{Abbot2012} also indicated that water worlds
might have a much narrower habitable zone than a planet with even a small land fraction. Moreover, these authors
suggested that a water world could ``self-arrest'' while undergoing a moist greenhouse from which the planet would
be left with partial ocean coverage and a benign climate. It is worth remarking that the importance of surface and
geologic effects on the water worlds is beyond the scope of this work. We just want to mention that the water worlds
represent a particular kind of exoplanets whose potential habitability should be studied with more detail.

We may wonder if the initial amount of mass in the HZ of the disk changes at the end of the simulations. Table
\ref{tab:5} shows the initial and final amounts of solid mass in the HZ for each density profile. Regarding $\gamma = 0.5$ 
and $\gamma = 1$ the values do not change significantly and this is because the gravitational interaccions between bodies in 
this region are weak and there is not a substantial mixing of solid material. Thus, embryos evolve very close to their initial 
positions and the initial mass in the HZ stays without significant changes. In contrast, the scenario is completely different for the 
third profile with $\gamma = 1.5$. As this profile is the more massive one in the inner zone, strong gravitational 
interactions are favored between bodies and some embryos migrate from the outer to the inner zone adding mass in the HZ.

\begin{table}[t!]
\caption{Initial and final amounts of solid mass in the HZ for each density profile.}
\begin{center}
\begin{tabular}{ccc}
\hline
\hline
$\gamma$  & Initial mass & Final mass \\
          & in the HZ    & in the HZ  \\
\hline
\hline
0.5       &  $0.104M_\oplus$   &   $0.101M_\oplus$ \\

1         &  $0.605M_\oplus$   &   $0.761M_\oplus$ \\

1.5       &  $1.423M_\oplus$   &   $2.867M_\oplus$ \\
\hline
\hline
\end{tabular}
\end{center}
\label{tab:5}
\end{table}

\begin{figure}[ht]
 \centering
 \includegraphics[angle=0, width= 0.45\textwidth]{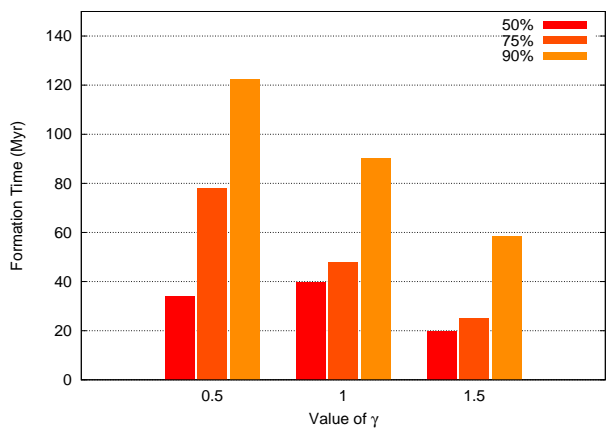}
 \caption{Timescales for planets in the HZ to reach a given fraction ($50\%$, $75\%$, or $90\%$)
of their final masses, as a function of the surface density profile $\gamma$. These values are
averages for all planets in the HZ of S1, S2 and S3 simulations.}
 \label{fig:Escalas}
\end{figure}

Figure \ref{fig:Escalas} shows the mean time values of the planets in the HZ to reach $50\%$, $75\%$ and $90\%$ of their final 
mass for the three values of $\gamma$. Planets in the HZ for $\gamma = 0.5$ form more slowly than those in the HZ for $\gamma = 1$ and 
$\gamma = 1.5$. Since this profile is the less massive in the inner zone of the disk, the gravitational interactions between bodies are
week. Thus, the acretion timescales for planets are longer than for the other profiles.

As we have already mentioned, regarding water delivery we can describe two scenarios within our simulations. 
Less massive disks in the inner zone, like those with $\gamma = 0.5$ and $\gamma = 1$, form 6 planets in the HZ 
with water contents ranging from $0.03\%$ to $9\%$ by mass. On the other hand, disks with $\gamma = 1.5$ which are 
more massive, form 4 planets in the HZ with water contents ranging from $4.51\%$ to $39.48\%$ by mass. In the first 
scenario planetesimals are mainly responsible for this water content, while in the second scenario we find planets
in which planetesimals are responsibles for their water content and planets in which embryos play the principal role.
In fact, for $\gamma = 1.5$, for 2 of the planets in the HZ that come from beyond the snow line, the
embryos are the responsibles of their water content and for the other 2 planets that come from the inner zone of the disk,
the planetesimals are mainly responsibles of their final water content. \citet{Morbidelli2000} 
showed that the bulk of Earth's water was acreted by a few asteroidal embryos from beyond 2 - 2.5~AU while 
\citet{Raymond2007b} proposed that terrestrial planets acrete a comparable amount of water in the form of a few 
water-rich embryos and millions of planetesimals. However, the simulations of \citet{Raymond2007b} show that the fraction 
of water delivered by planetesimals is much larger than the one delivered by embryos \footnote{It is worth noting that a 
comparison between \citet{Morbidelli2000}, \citet{Raymond2007b} and this work should be taken carefully since the scenarios and
their treatments are quite different, mainly because our scenario does not include a giant planet.}. Despite the architecture of 
our planetary systems is very different from the one studied by \citet{Morbidelli2000} and \citet{Raymond2007b} 
since we do not consider the existence of giant gaseous planets, the results of the first scenario of our 
simulations present the same trend as \citet{Raymond2007b}. Nevertheless, when the mass of the inner part of the 
disk increases, we find that some of our results are consistent with those of \citet{Morbidelli2000} and some are
consistent with those of \citet{Raymond2007b}.

\citet{Raymond2005} analyzed the terrestrial planet formation in disks with varying surface density profiles. To do 
this, they considered that the surface density varies as $r^{-\gamma}$, and assumed three different values for $\gamma$: 0.5, 1.5 
and 2.5. It is worth emphasizing that these authors did not take into account an increase in surface density due to the condensation 
of water beyond the snow line. On the other hand, \citet{Raymond2005} considered a fixed mass of material equals to $10M_\oplus$ 
between 0.5~AU and 5~AU for each distribution, and besides they assumed that the disk mass is dominated by embryos, which have 
swept up the mass in their corresponding feeding zones. Finally, these authors developed the simulations including a Jupiter-mass 
giant planet at 5.5~AU. In this scenario, the individual masses of the planetary embryos in the outer region of the system are 
larger for smaller values of gama. The larger the mass of the embryos in the outer region, the more significant the scattering 
of water-rich material on the entire system. For this reason, \citet{Raymond2005} found a more efficient water delivery for 
$\gamma = 0.5$ than for $\gamma = 1.5$.

In our simulations, we assume surface density distributions of the form $r^{-\gamma}$, and explore three different values for $\gamma$ 
of 0.5, 1 and 1.5. Unlike \citet{Raymond2005}, we assume a disk with a fixed mass of $0.03M_\odot$, which leads to a total mass of 
solids between 0.5~AU and 5~AU of $3.21M_\oplus$, $7.92M_\oplus$, and $13.66M_\oplus$, for $\gamma = 0.5$, $1$, and $1.5$, respectively. 
Thus, in our work, the individual masses of the planetary embryos in the outer region of the system are larger for higher values of $\gamma$. 
From this, the distribution with $\gamma = 1.5$ leads to a more significant scattering of water-rich material associated to the 
outer region of the system. For this reason, our results suggest that the water delivery is more efficient for $\gamma = 1.5$ than 
for $\gamma = 0.5$. It is worth noting that our results are consistent with those obtained by \citet{Raymond2005}, since in both 
studies, the water delivery is more efficient in systems that contain the more massive embryos in the outer region.

To determine the final water content of the resulting planets of our simulations, we adopted an initial
distribution of water content that is based on data for primitive meteorites from \citet{Abe2000}.
We analyzed the sensitivity of our results to the initial distribution of water assumed for the protoplanetary
disk. To do this, we considered a simple prescription for assigning initial water contents to embryos and planetesimals
as a function of their radial distances. In fact, we assumed that bodies inside 2.7 AU do not have water while
bodies beyond 2.7 AU contain 50 \% water by mass. From this new initial distribution, we did not find relevant
changes in the final water contents of the resulting planets of our simulations. This result confirms that the
water delivery to the planets located at the HZ is provided primarily by embryos and planetesimals starting the simulation
beyond the snow line. In fact, the initial water content of the bodies located inside the snow line does not lead
to relevant changes in our results.

By analyzing the masses and the water contents of the planets formed in the HZ of our simulations, we conclude
that the surface density profiles with $\gamma =$ 1 and 1.5 produce planetary systems of special astrobiological
interest from a low-mass disk. It results very interesting to discuss if planets analogous to those formed in the HZ
of such systems can be discovered with the current detection techniques. The NASA Kepler mission\footnote{http://kepler.nasa.gov/}
was developed with the main purpose of detecting Earth-size planets in the HZ of solar-like stars \citep{Koch2010}.
To date, this mission has discovered 167 confirmed planets and over 3538 unconfirmed planet candidates. The Kepler-37
system hosts the smallest planet yet found around a star similar to our Sun \citep{Barclay2013a}. This planet, which is called
Kepler-37b, is significantly smaller than Mercury and is the innermost of the three planets of the system
at 0.1 AU. On the other hand, the smallest habitable zone planets discovered to date by the Kepler mission
are Kepler-62e, 62f \citep{Borucki2013} and 69c \citep{Barclay2013b} with 1.61, 1.41, and 1.71 Earth radii, respectively.
The potentially habitable planets formed in all our simulations have sizes ranging from 0.38 to 1.6 Earth radii, assuming
physical densities of 3 g cm$^{-3}$. Thus, the Kepler mission would seem to be able to detect the potentially habitable planets
of our simulations very soon. However, in May 2013, Kepler spacecraft lost the second of four gyroscope-like reaction wheels,
ending new data collection for the original mission. Currently, the Kepler mission has assumed a new concept, dubbed K2,
in order to continue with the search of other worlds. A decision about it is expected by the end of 2013. 
Future missions, such as
PLAnetary Transits and Oscillations of stars (PLATO 2.0; \citealt{Rauer2013}), will play a significant role in the detection 
and characterization of terrestrial planets in the habitable zone around solar-like stars during the next decade. 
In fact, the primary goal of PLATO is to assemble the first catalogue of confirmed planets in habitable zones with known 
mean densities, compositions, and evolutionary stages. This mission will play a key role in determining how common worlds 
like ours are in the Universe as well as how suitable they are for the development and maintenance of life.  

Beyond the detection of planets in the HZ, we can ask if it is possible to distinguish the planetary systems of interest
obtained in our simulations. The gravitational microlensing technique will probably play a significant role in the detection
of planetary systems similar to those obtained in our work. In fact, unlike other techniques such as the transit method or
radial velocities, the microlensing technique is sensitive to planets on wide orbits around the snow line of the system
\citep{Gaudi2012}. Currently, the main microlensing surveys for exoplanets are the Optical Gravitational Lensing Experiment
(OGLE; \citealt{Udalski2003}) and the Microlensing Observations in Astrophysics (MOA; \citealt{Sako2008}). To date, a total number of
26 planets have been detected by such surveys. The less massive planets discovered to date by the microlensing technique are
MOA-2007-BLG-192-L b \citep{Bennett2008} and OGLE-05-390L b \citep{Beaulieu2006} with 3.3$^{+4.9}_{-1.6}$ $M_{\oplus}$ and
5.5$^{+5.5}_{-2.7}$ $M_{\oplus}$, respectively. It is worth remarking that MOA-2007-BLG-192-L b and OGLE-05-390L b orbit
stars with 0.06$^{+0.028}_{-0.021}$ $M_{\odot}$ and 0.22$^{+0.21}_{-0.11}$ $M_{\odot}$, respectively. The lowest mass exoplanet
found to date orbiting a Sun-like star is OGLE-2012-BLG-0026L b \citep{Han2013}. This planet is located at $\sim$ 3.8 AU from the
central star and has 0.11 $\pm$ 0.02 $M_{\textrm{J}}$, where $M_{\textrm{J}}$ represents a Jupiter mass.

The planets formed in our simulations that might be found by the microlensing technique are significantly less massive that those
detected to date by such technique. In fact, for $\gamma =$ 0.5, our simulations formed planets of $\sim$ 0.5 $M_{\oplus}$ between
2.1 AU and 4.3 AU. For $\gamma =$ 1, the resulting system shows planets with masses ranging from 1.4 $M_{\oplus}$ to 1.9 $M_{\oplus}$
between 1.7 AU and 3.2 AU. Finally, for $\gamma =$ 1.5, our simulations produced planets with masses of 2.2 $M_{\oplus}$ to 3.1
$M_{\oplus}$ between 2.7 AU and 3.4 AU. As we have already mentioned in the last paragraph, the current microlensing surveys
have not detected yet planets analogous to those formed on wide orbits in our simulations. However, future surveys such as
the Korean Microlensing Telescope Network (KMTNet; \citealt{Poteet2012}) and the Wide-Field InfraRed Survey Telescope  
(WFIRST; \citealt{Green2011}) will play a relevant role in the search of exoplanets by the microlensing technique. On the one hand, 
KMTNet is a
ground-based project with plans to start operations in 2015. On the other hand, WFIRST is a space-based project which could be ready
for launch in 2020. In particular, the main goal of WFIRST is to detect via microlensing planets with masses
$\gtrsim$ 0.1 $M_{\oplus}$ and separations $\gtrsim$ 0.5 AU, including free-floating planets. Thus, we think that planetary systems
similar to those formed in the present work should be detected by microlensing techniques within this decade.

The main result of this study suggests that the planetary systems without gas giants that harbor 1.4-3.1 $M_{\oplus}$ planets on wide
orbits around Sun-like stars are very interesting from an astrobiological point of view. This work complements that developed by
\citet{deElia2013}, which indicates that systems without gas giants that harbor super-Earths or Neptune-mass planets on
wide orbits around solar-type stars are of astrobiological interest. These theoretical works offer a relevant contribution for current
and future observational surveys since they allow to determine planetary systems of special interest.
\begin{acknowledgements}
    We are grateful to Pablo J. Santamar\'{\i}a who provided us with the numerical tools necessary to study the collisional history 
and water accretion in planets in the HZ. We also want to thank Juan P. Calder\'on who kindly helped us to improve the plots of 
the time evolution of all our simulations. 
\end{acknowledgements}


\bibliographystyle{aa} 


\end{document}